\documentclass[preprint2]{aastex}

\shorttitle{Pixel Analysis of NGC 5194}

\shortauthors{Lee et al.}
\def\simlt{\lower.5ex\hbox{$\; \buildrel < \over \sim \;$}}
\def\simgt{\lower.5ex\hbox{$\; \buildrel > \over \sim \;$}}

\begin{document}

\title{\emph{HUBBLE SPACE TELESCOPE} PIXEL ANALYSIS OF THE INTERACTING FACE-ON SPIRAL GALAXY NGC 5194 (M51A)}

\author{Joon Hyeop Lee $^1$, Sang Chul Kim $^1$, Hong Soo Park $^{1,2}$, Chang Hee Ree $^1$, Jaemann Kyeong $^1$, Jiwon Chung $^{1,3}$}
\affil{$^1$ Korea Astronomy and Space Science Institute, Daejeon 305-348, Republic of Korea\\
$^2$ Astronomy Program, Department of Physics and Astronomy, Seoul National University, Seoul 151-742, Republic of Korea\\
$^3$ Department of Astronomy and Space Science, Chungnam National University, Daejeon 305-764, Republic of Korea}

\email{jhl@kasi.re.kr, sckim@kasi.re.kr, hspark@astro.snu.ac.kr, chr@kasi.re.kr, jman@kasi.re.kr, jiwon@kasi.re.kr}

\begin{abstract}
A pixel analysis is carried out on the interacting face-on spiral galaxy NGC 5194 (M51A), using the {\it HST}/ACS images in the F435W, F555W and F814W ($BVI$) bands. After $4\times4$ binning of the {\it HST}/ACS images to secure a sufficient signal-to-noise ratio for each pixel, we derive several quantities describing the pixel color-magnitude diagram (pCMD) of NGC 5194: blue/red color cut, red pixel sequence parameters, blue pixel sequence parameters and blue-to-red pixel ratio.
The red sequence pixels are mostly older than 1 Gyr, while the blue sequence pixels are mostly younger than 1 Gyr, in their luminosity-weighted mean stellar ages. The color variation in the red pixel sequence from $V=20$ mag arcsec$^{-2}$ to $V=17$ mag arcsec$^{-2}$ corresponds to a metallicity variation of $\Delta$[Fe/H] $\sim2$ or an optical depth variation of $\Delta\tau_V\sim4$ by dust, but the actual sequence is thought to originate from the combination of those two effects. At $V<20$ mag arcsec$^{-2}$, the color variation in the blue pixel sequence corresponds to an age variation from 5 Myr to 300 Myr under the assumption of solar metallicity and $\tau_V=1$.
To investigate the spatial distributions of stellar populations, we divide pixel stellar populations using the pixel color-color diagram and population synthesis models. As a result, we find that the pixel population distributions across the spiral arms agree with a compressing process by spiral density waves: dense dust $\rightarrow$ newly-formed stars. The tidal interaction between NGC 5194 and NGC 5195 appears to enhance the star formation at the tidal bridge connecting the two galaxies.
We find that the pixels corresponding to the central active galactic nucleus (AGN) area of NGC 5194 show a tight sequence at the bright-end of the pCMD, which are in the region of $R\sim100$ pc and may be a photometric indicator of AGN properties.
\end{abstract}


\section{INTRODUCTION}

NGC 5194 is one of the nearest galaxies with grand design spiral arms, also known as M51A\footnote{The name `M51' often designates the interacting system of NGC 5194 + 5195, while it sometimes indicates NGC 5194 only. In this paper, to avoid the terminological confusion, we use the name `M51' only when we mean the NGC 5194 + 5195 system. } or the Whirlpool galaxy. NGC 5194 is a very interesting object not only because we can inspect its spiral arm structures with close and face-on views, but also because NGC 5194 is known to be interacting with its companion galaxy NGC 5195. The morphological type of NGC 5194 is Sbc \citep{dev91} and that of NGC 5195 is SB0 \citep{san87}. Due to such interesting properties, the entire area of the interacting system NGC 5194 + 5195 was covered using the {\it Hubble Space Telescope} ({\it HST}) and Advanced Camera for Surveys (ACS) with very high resolution by the Hubble Heritage Team \citep{mut05}, despite its large angular size ($\sim9'$).

Owing to its scientific attraction and high-quality data, NGC 5194 has been investigated by many researchers.
One of the active studies on NGC 5194 using the {\it HST} data is the search and investigation of star clusters in it. Until now, about 2000 star clusters in M51 have been found and their physical properties, such as age and mass, were derived, revealing that the star cluster formation rate increased significantly during the period of $100-250$ Myr ago, which is consistent with the epoch of the dynamical encounters of the two galaxies \citep[e.g.][]{lee05,hwa08,hwa10}.
Bright stars are resolved as point sources in NGC 5194, like the star clusters. \citet{kal10} presented a photometric study of the stellar associations across the entire disk of NGC 5194, determining the age, size, mass, metallicity and dust content of each association as a function of location.
In addition to the stellar properties, the collective properties of the H{\scriptsize II} regions in M51 were also investigated using the \emph{HST} $H{\alpha}$-band data in \citet{lee11}. They found that the $H{\alpha}$ luminosity functions have different slopes for different parts in M51, which implies that the H{\scriptsize II} regions in the interarm region are relatively older than those in the other parts of M51.

The effect of interaction on the properties of NGC 5194 is a topic studied by many researchers.
\citet{dur03} investigated the kinematics of planetary nebulae in M51's tidal debris, finding that the tidal debris consists of two discrete structures: NGC 5195's own tidal tail and diffuse material stripped from NGC 5194, which is consistent with the `single-passage' model for the encounter with a 2:1 mass ratio.
\citet{mei08} tested whether the strong spiral structure is transient (i.e., interaction-driven) or steady, based on the radial dependence of the pattern speed, showing two distinct pattern speeds inside 4 kpc at their derived major axis, which are indicative of a long-lasting spiral structure.

The studies on the nuclear region of NGC 5194 are also active.
\citet{gri97} analyzed the central region of NGC 5194 imaged with the {\it HST}, finding a complicated distribution of dust lanes. Those dust lanes are roughly aligned with the major axis of the bar and may be transporting gas to the AGN in the nucleus.
Later, a Chandra observation on the AGN and nuclear outflows of NGC 5194 was carried out \citep{ter01} and showed a low-luminosity Seyfert 2 nucleus, southern extra-nuclear cloud and northern loop. The morphology of the extended emission is very similar to radio continuum and optical emission-line images.
\citet{lam02} inspected the ongoing massive star formation in the bulge of NGC 5194 using the {\it HST}/WFPC2 observation data and revealed that the bulge around the nucleus is dominated by a smooth reddish background population with overimposed dust lanes. They found 30 bright point-like sources in the bulge within 110-350 pc of the nucleus, which are probably isolated massive stars.
\citet{mat04,mat07} studied the jet-disturbed molecular gas near the NGC 5194 AGN, which showed the presence of a dense circumnuclear-rotating disk, which may be a reservoir that fuels the active nucleus and obscures it from direct optical view.

Those various studies revealed many interesting properties of NGC 5194, but there remains a useful analysis method not previously tried for NGC 5194: {\it pixel analysis}.
Recently, pixel analysis methods started to be used for investigation of various galaxies, thanks to the observational facilities and techniques that provide high spatial resolution and high signal-to-noise ratio (S/N) even for individual pixels.
One of the well-known pixel analysis methods is the {\it pixel-z} technique established by \citet{con03} to investigate the star formation history of galaxies measured from individual pixels.
\citet{joh05} carried out pixel-by-pixel SED fitting for a galaxy hosting a compact steep-spectrum radio source MRC B1221$-$423, based on $BVRIK$ imaging and spectroscopic observations.
More recently, the pixel-z technique was applied for 44,964 galaxies to derive their ages, star formation rates, dust obscurations, metallicities, and morphology -- density relations \citep{wel08,wel09}.
\citet{wij10} extracted the physical properties of the SDSS galaxies with the pixel-z method, relating the spatial distribution of the physical properties with the morphological properties of galaxies.

The pixel-z method is powerful, but it is based on SED fitting, which depends on SED templates and fitting algorithms and requires images in many (at least four) bands. A simpler way is the pixel color-magnitude or color-color diagram analysis, which was tried a long time ago for estimating stellar populations in unresolved galaxies \citep[e.g.][]{bot86}.
\citet{kas03} studied the stellar populations of merging galaxies NGC 4038/4039 in the $BVK$ bands, deriving pixel-by-pixel maps of the distributions of stellar populations and dust extinction. From the pixel color-color diagrams (pCCDs) of those galaxies and population synthesis models, they estimated the approximate age and reddening of each pixel in those galaxies.
Similarly, \citet{deg03} analyzed the pixel color-magnitude diagram (pCMD) and pCCD, to study the star formation histories of the Mice (NGC 4676) and Tadpole (UGC 10214) interacting galaxies. The spatial distributions of pixels in different domains of the pCMD and pCCD were investigated and revealed that galaxy interactions significantly affect stellar populations.

More recently, a systematic comparison of the pCMD features between galaxies with different morphologies was carried out by \citet{lan07} for a large number of galaxies. They studied 69 nearby galaxies chosen to span a wide range of Hubble types using their pCMDs and found that early-type galaxies exhibit clear \emph{prime sequences} but some of them have dust-originated \emph{red hook} features. They also found that lenticular galaxies show pCMD features similar to those of elliptical galaxies but tend to have larger color dispersions and that some pCMDs of face-on spirals show \emph{inverse-L} features.
In addition, they developed quantitative methods to characterize the pCMDs, such as the blue-to-red light ratio and total color dispersion.
\citet{lan07} suggested a leading concept in the methodology of pCMD analysis: the \emph{quantification} of pCMD features. By quantifying the various features in pCMDs, it is possible to compare the pCMDs systematically between different galaxies, which is a totally new methodology in galaxy studies using simple (but highly-resolved) image data.
However, since the quantified parameters in \citet{lan07} are so simple they are not enough to describe the various features of pCMDs, that is, the pCMD analysis method is not perfectly established yet.

In this paper, we present a pixel analysis of the interacting face-on spiral galaxy, NGC 5194.
The main goals of this paper are 1) to improve the understanding of NGC 5194 properties using pixel analysis methods and 2) to establish the pCMD analysis methods for future photometric studies of galaxies. For the second goal, NGC 5194 (or M51) is an ideal target, because it is a very nearby face-on galaxy, and the entire area was covered by the {\it HST} with very high spatial resolution. With the high-resolution image data, how the pCMD features vary as a function of spatial resolution can be tested.

The outline of this paper is as follows. Section 2 describes the data and \S3 explains the pixel binning and analysis area. The results of the pixel analysis are shown in \S4 and discussed in \S5. Section 6 gives the conclusion. Throughout this paper, we adopt
the cosmological parameters: $h=0.7$, $\Omega_{\Lambda}=0.7$, and
$\Omega_{M}=0.3$.

\section{DATA}

All the work in this paper was carried out using the data set obtained by the Hubble Heritage Team. They observed M51 using the {\it HST} Advanced Camera for Surveys (ACS) with F435W, F555W, F814W and F658N filters as part of {\it HST} program 10452. The observation was completed in January 2005 and the data were publicly released in April 2005, covering about a $6.8' \times 10.5'$ field centered on M51. The accumulated exposure times are 2720, 1360 and 1360 seconds in F435W, F555W and F814W, respectively. The basic data processing, multi-drizzling and image combination were done by the Space Telescope Science Institute (STScI) before the data release. More details on the observation and data reduction are available in \citet{mut05}.
In this paper, we used F435W, F555W and F814W (hereafter, we refer to these filters simply as $B$, $V$ and $I$, respectively) images.
Pixel surface brightness and its photometric error were derived using the science and weight images\footnote{The M51 Hubble Heritage Program data and related information can be retrieved from the following webpage: http://archive.stsci.edu/prepds/m51/index.html}. Based on the instruction of the {\it HST} Data Handbook for ACS (version 5.0), we derived the following equations for pixel AB magnitude and photometric error in each band:
\begin{equation}
m_F = -2.5 \log P_F + Z_F,
\end{equation}
\begin{equation}
e_F = 2.5 \log e \times (P_F \times W_F)^{-0.5},
\end{equation}
where $m_F$, $e_F$, $P_F$ and $W_F$ are the pixel AB magnitude, photometric error, science-image pixel value and weight-image pixel value in the $F$ band, respectively. $Z_F$ is the magnitude zero point in the $F$ band, which is 25.6732, 25.7349 and 25.9366 for the $B$, $V$ and $I$ bands, respectively. Surface brightness is defined as AB magnitude per unit angular area (1 arcsec$^{2}$).

The distance to NGC 5194 was estimated by \citet{fel97} using the planetary nebula luminosity function, which we adopted in this paper: $8.4\pm0.6$ Mpc ($m-M=29.62\pm0.15$). At this distance, the linear scale is 40 parsecs per arcsecond, which corresponds to 2.0 parsecs per pixel for an {\it HST}/ACS pixel size of $0.05''$. The foreground reddening toward NGC 5194 is $E(B-V)=0.035$, with the extinction in each band: $A_B=0.150$, $A_V=0.115$ and $A_I=0.067$ mag, which was calculated using the reddening map of \citet{sch98} and the reddening law of \citet{car89}. All magnitudes, surface brightnesses and colors in this paper have been corrected for the foreground extinction.

\section{PIXEL BINNING AND ANALYSIS AREA}

\begin{figure}[!t]
\plotone{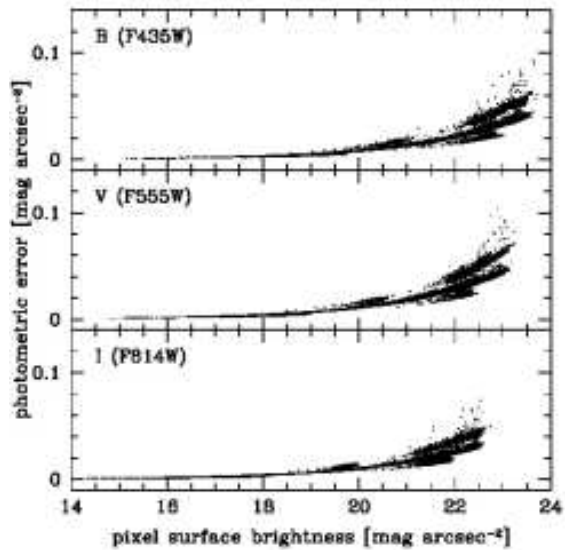}
\caption{ Binned-pixel (1 binned-pixel = $4 \times 4$ pixels = $0.2 \times 0.2$  arcsec$^2$) surface brightnesses and photometric errors in the $B$ (F435W), $V$ (F555W) and $I$ (F814W) bands. Only $1\%$ (selected randomly) of the entire pixels are displayed.
\label{photo}}
\end{figure}

The angular pixel scale of the {\it HST}/ACS is $0.05''$, but the typical full width at half maximum (FWHM) of a point source in the {\it HST}/ACS is about $0.1''$.
Thus, to ensure that each binned pixel is statistically independent of the surrounding pixels in the pixel analysis, it is necessary to bin the pixels by a factor of at least 2. In this paper, we binned the pixels by a factor of 4 ($4\times4$ binning) to improve the reliability of the pixel values (that is, to improve the S/N for each pixel).
Fig.~\ref{photo} displays the photometric error for the binned-pixel surface brightness, showing that even the faintest pixel has relatively good photometric quality (photometric error $\simlt0.1$ mag arcsec$^{-2}$).
In Fig.~\ref{photo}, some discrete features are shown at the faint end, because the M51 images are drizzled ones. The Hubble Heritage Team provides a single drizzled image in each band, but it is actually a composition of multiple images, that is, the exposure time is not homogeneous for every pixel but different and discontinuous.

\begin{figure}[!t]
\plotone{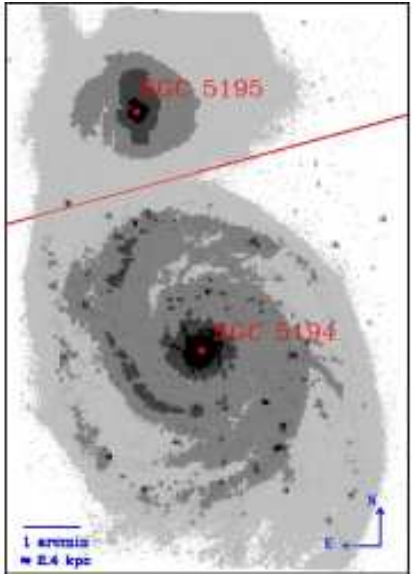}
\caption{ Area definition of M51 (NGC 5194 + 5195). The darker pixels indicate the brighter surface brightnesses. The solid line is the 7:3 division between NGC 5194 and NGC 5195 centers. In this paper, the analysis is carried out for the area below the solid line (NGC 5194-side).
\label{arealimit}}
\end{figure}

We analyze NGC 5194 only, because NGC 5195 has much higher level of internal dust attenuation than NGC 5194 does, which makes the pixel properties of NGC 5195 very different from those of NGC 5194.
To define the NGC 5194 area, we draw a line (hereafter, the `division line') that is vertical to the line connecting the centers of NGC 5194 and NGC 5195. The ratio between the distances of the division line to the NGC 5194 center and to the NGC 5195 center is manually selected to be 7:3, as shown in Fig.~\ref{arealimit}. The central coordinate (J2000) of NGC 5194 is RA = 13h 29m 52.72s and Dec = +47d 11m 43.4s, while that of NGC 5195 is RA = 13h 29m 59.54s and Dec = +47d 15m 58.3s, which are directly estimated using the {\it HST}/ACS $V$-band drizzled image.
The pixel analysis in this paper is carried out for the `NGC 5194 area' defined in Fig.~\ref{arealimit} (below the solid line).

\section{RESULTS}\label{result}

\subsection{pCMD}

\subsubsection{Color distribution as a function of surface brightness}\label{cdist}

\begin{figure}[!t]
\plotone{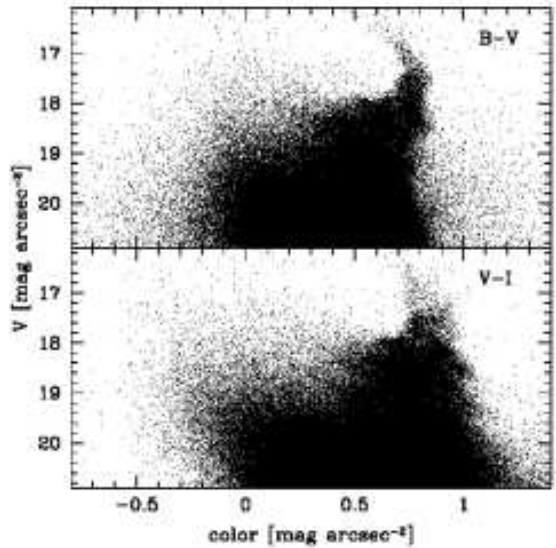}
\caption{ Pixel color versus magnitude diagrams (pCMDs) of NGC 5194 with $4\times4$ binning: $V$ versus $B-V$ (upper panel) and $V$ versus $V-I$ (lower panel).
\label{allcmd}}
\end{figure}

We investigated the pixel color-magnitude relation of NGC 5194 using the pCMDs with $4\times4$ binning, as shown in Fig.~\ref{allcmd}.
In those pCMDs, most pixels form a sequence with large dispersion, in which brighter pixels tend to be redder. In that `sequence', faint pixels have a large color dispersion, while the colors of bright pixels have a narrow range.
However, there are some bright ($V<19$ mag arcsec$^{-2}$) \footnote{Here, `$V$' actually indicates the surface brightness in the $V$ band (i.e., $\mu_V$). Throughout this paper, $\mu_V$ is simply denoted as $V$ for convenience. } pixels with blue colors (e.g., $V-I<0.4$), which have a very small number fraction compared to the bright pixels in the `sequence' (hereafter, red pixel sequence). Those bright blue pixels seem to form another sequence with a large color dispersion rather than being continuously dispersed from the red pixel sequence. The two tentative (red and blue) pixel sequences are obviously distinct for bright pixels, whereas they are less distinct for fainter pixels. The overall shape of those sequences is like two branches with different thicknesses, stemming from a single root.

\begin{figure}[!t]
\plotone{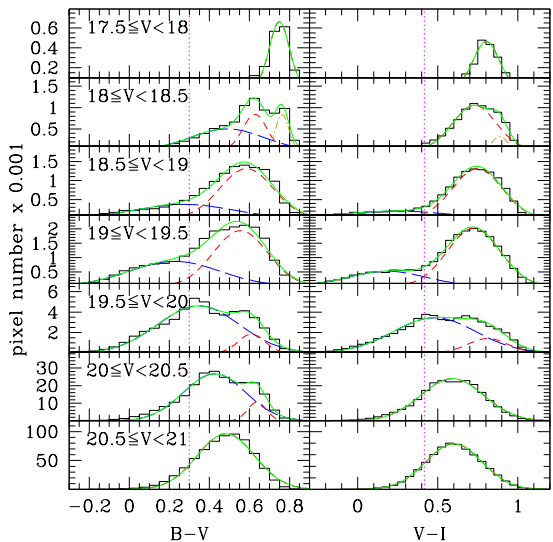}
\caption{ Color distributions as a function of pixel surface brightness with Gaussian fits. The dashed curves are individual Gaussians and the solid curves are the summed forms. The vertical dotted-lines show the $B-V=0.299$ and $V-I=0.416$ color cuts, which are the `red Gaussian peak color' $-$ 2 $\times$ `red Gaussian dispersion' values ($x_2-2\sigma_2$; see Table~\ref{gauss}) at $18.5\le V<19.5$ mag arcsec$^{-2}$.
\label{colfit}}
\end{figure}

For a quantitative analysis of the pCMD features, we investigated the color distribution as a function of pixel surface brightness. In Fig.~\ref{colfit}, the pixel color distribution is presented for every 0.5 mag arcsec$^{-2}$ interval between $17.5\le V<21.0$ mag arcsec$^{-2}$ and fit using single, double or triple Gaussian functions. The Gaussian fit parameters are summarized in Table~\ref{gauss}. The color distributions for the brightest pixels and for the faintest pixels in Fig.~\ref{colfit} are fit well by single Gaussians, but the intermediate brightness ($18\le V<20.5$ mag arcsec$^{-2}$) pixels show mostly bimodal color distributions.
It is noted that the pixels at $18.0\le V<18.5$ mag arcsec$^{-2}$ show a very red component (the third Gaussian), which mostly corresponds to the pixels with heavy dust extinction in the bulge.
Double Gaussians are most largely separated at $18.5\le V<19.5$ mag arcsec$^{-2}$, where the double Gaussian peak$\pm$dispersion values are $B-V=0.247\pm0.237$ and $0.568\pm0.135$; and $V-I=0.212\pm0.279$ and $0.736\pm0.160$. At $18.5\le V<19.5$ mag arcsec$^{-2}$, the blue-to-red pixel number ratio (blue/red ratio) estimated using the blue and red Gaussians (that is, the Gaussian area ratio) is 0.689 for $B-V$ and 0.381 for $V-I$.

The double Gaussian fitting is a reasonable method to segregate blue/red pixels and to estimate the blue/red ratio, but it has two problems. First, as shown in Fig.~\ref{colfit}, double Gaussians are clearly separated only at a narrow range of surface brightness, and thus it is difficult to apply it to all the pixels. Since the pixels at $18.5\le V<19.5$ mag arcsec$^{-2}$ are a small part of the entire pixel set, those pixels may not necessarily represent the total pixel properties. At the surface brightness range where the color separation is not clear (e.g., $19.5\le V<20$ mag arcsec$^{-2}$), the double Gaussian fit results significantly depend on the initial values, making their reliability low.
Second, it is difficult to apply the double Gaussian fitting method to pCMDs with low pixel resolutions, since it requires at least hundreds of pixels at a given surface brightness interval for statistical reliability. Thus, this method is limited to nearby galaxies imaged using good facilities and thus applicable to a very small sample of galaxies.

For those reasons, we devised a simple method to quantify the shape of pCMDs. In this method, we divide the pixels into blue and red ones with fixed color cuts, which are based on the double Gaussian fits at $18.5\le V<19.5$ mag arcsec$^{-2}$.  As the color cuts, we selected the `red Gaussian peak color' $-$ 2 $\times$ `red Gaussian dispersion' values ($x_2-2\sigma_2$; see Table~\ref{gauss}) at $18.5\le V<19.5$ mag arcsec$^{-2}$, which are $B-V=0.299$ and $V-I=0.416$. We based the color cuts on the red Gaussian not the blue Gaussian, because (1) red pixel sequences are commonly found both in early-type and late-type galaxies and (2) blue pixel sequences are too sensitive to short-timescale variations like recent star formation.
The vertical lines in Fig.~\ref{colfit} show those color boundaries, dividing appropriately the blue and red pixels.

\begin{figure}[!t]
\plottwo{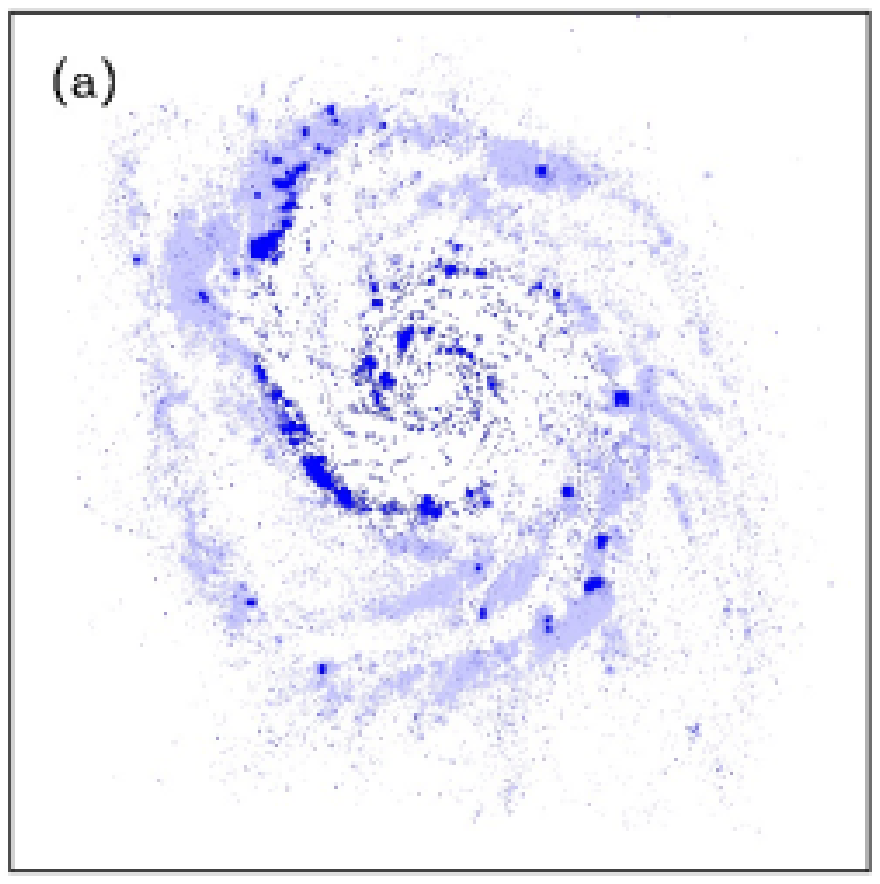}{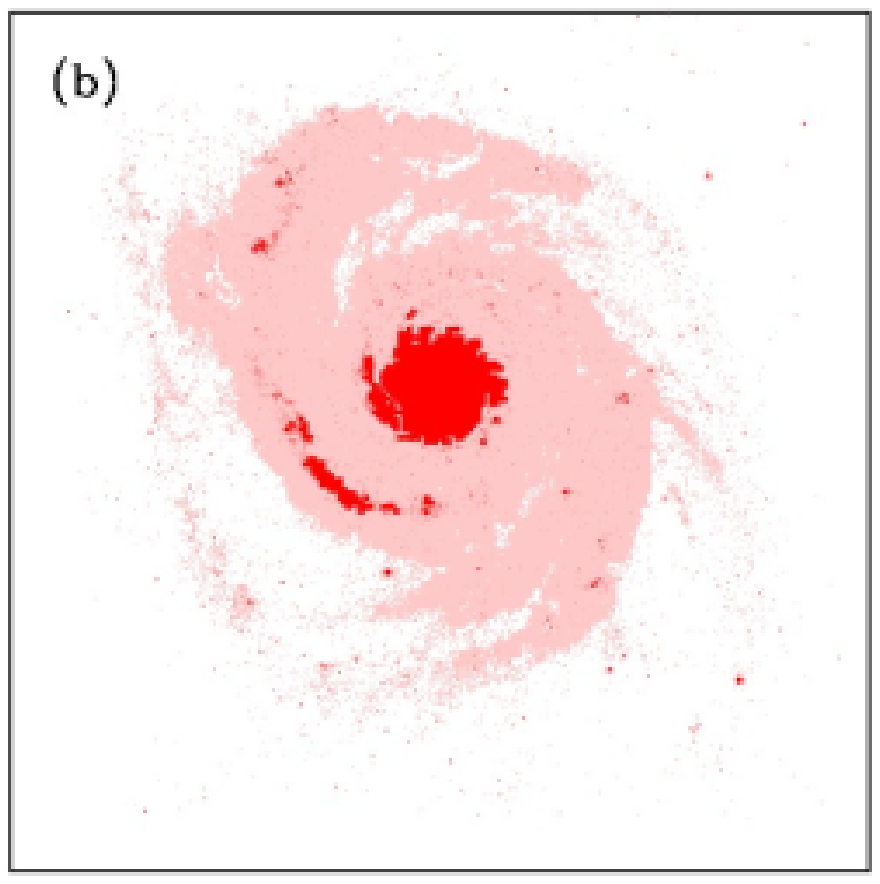}
\caption{ Spatial distributions of pixels divided by the color cut: (a) $V-I<0.416$ and (b) $V-I\ge0.416$, for $V<20$ mag arcsec$^{-2}$ (dark dots) and for $20\le V<21$ mag arcsec$^{-2}$ (light dots).
\label{cmap}}
\end{figure}

Fig.~\ref{cmap} displays the spatial distributions of the blue and red pixels at $V<21$ mag arcsec$^{-2}$. It is obvious that blue pixels are biased to disk and spiral arms, while red pixels are more concentrated on the bulge area than blue pixels. However, it is noted that a large number of red pixels are also found along the spiral arms. More detailed analysis on the pixel stellar populations and their spatial distribution will be shown in \S\ref{svsp}. The pixels at $V<20$ mag arcsec$^{-2}$ reveal the spiral-arm patterns clearly, but they miss a large fraction of pixels in NGC 5194. Thus, we use the pixels at $V<20$ mag arcsec$^{-2}$ for the simple inspection of pixel distribution and spiral arm structure definition (e.g., \S\ref{armdef}), while we use the pixels at $V<21$ mag arcsec$^{-2}$ for the statistical analysis of pixel populations (e.g., \S\ref{popvartext}).

\begin{figure*}[!t]
\epsscale{2}
\plotone{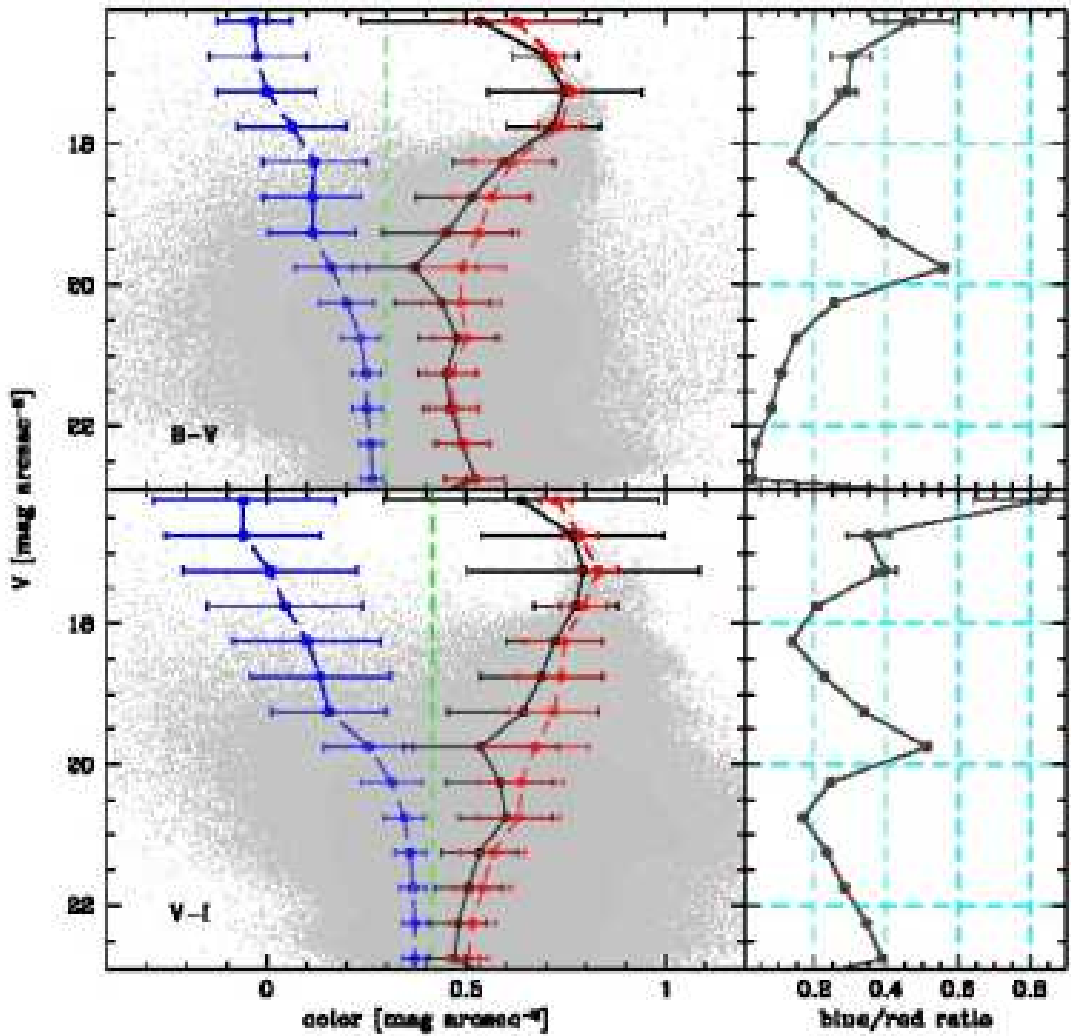}
\caption{ {\it Left panels}: Blue (long-dashed curves), red (short-dashed curves) and total (solid curves) pixel sequences in the $4\times4$ binned pCMDs. The vertical dashed-line is the color division: $B-V=0.299$ and $V-I=0.416$. Open circles show the median colors at given $V$ surface brightness ranges for each sequence and the errorbars indicate the sample inter-quartile ranges (SIQRs). {\it Right panels}: Blue-to-red pixel number ratio with Poisson error.
\label{pcmd}}
\epsscale{1}
\end{figure*}

Fig.~\ref{pcmd} shows some statistics of the pCMDs divided into blue and red pixels. As already described qualitatively, the blue pixels and red pixels form two distinct sequences: the brighter pixels tend to be bluer in the blue pixel sequence, while they tend to be redder in the red pixel  sequence. Another noticeable result is that the blue/red ratio shows a peak at $19.5\le V<20.0$ mag arcsec$^{-2}$ both in the $B-V$ and in the $V-I$ colors. Since the spiral-arm patterns (that is, the spatial distributions of the high-density area of blue stars) are seen most obviously at $V<20$ mag arcsec$^{-2}$, the blue/red ratio peak seems to be mainly due to the young stellar populations in the spiral arms. The blue/red ratio shows a bottom at $18.0\le V<18.5$ mag arcsec$^{-2}$ both in the $B-V$ and $V-I$ colors and another bottom at $20.5\le V<21.0$ mag arcsec$^{-2}$ in the $V-I$ color. The parameters describing the shapes of the pCMDs are listed in Table~\ref{pcmdinfo}.

\begin{figure}
\plotone{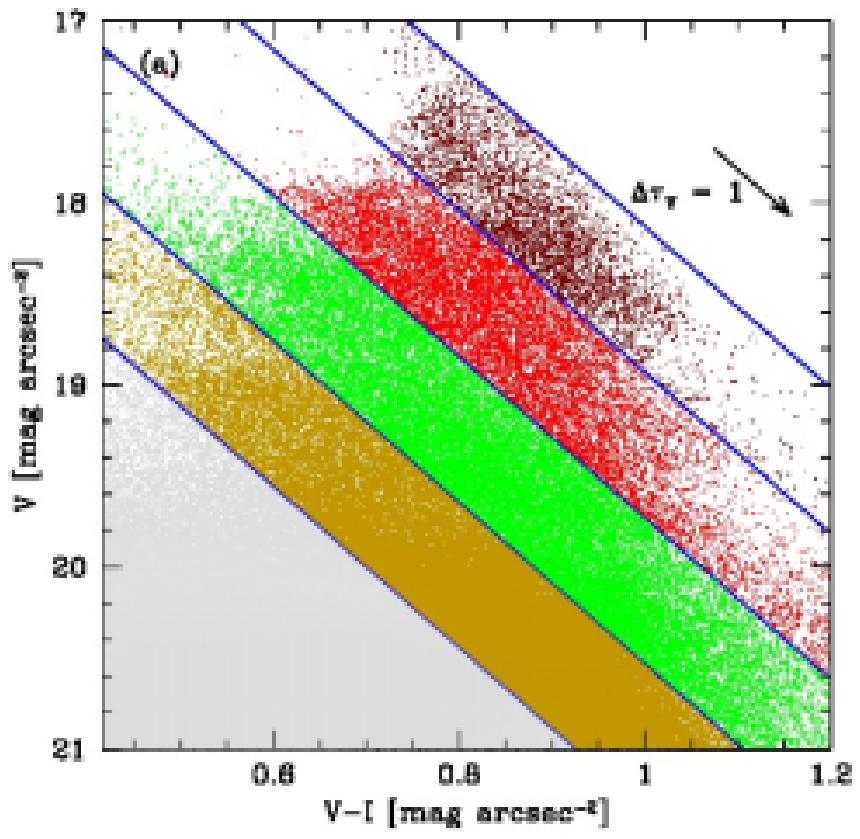}
\plotone{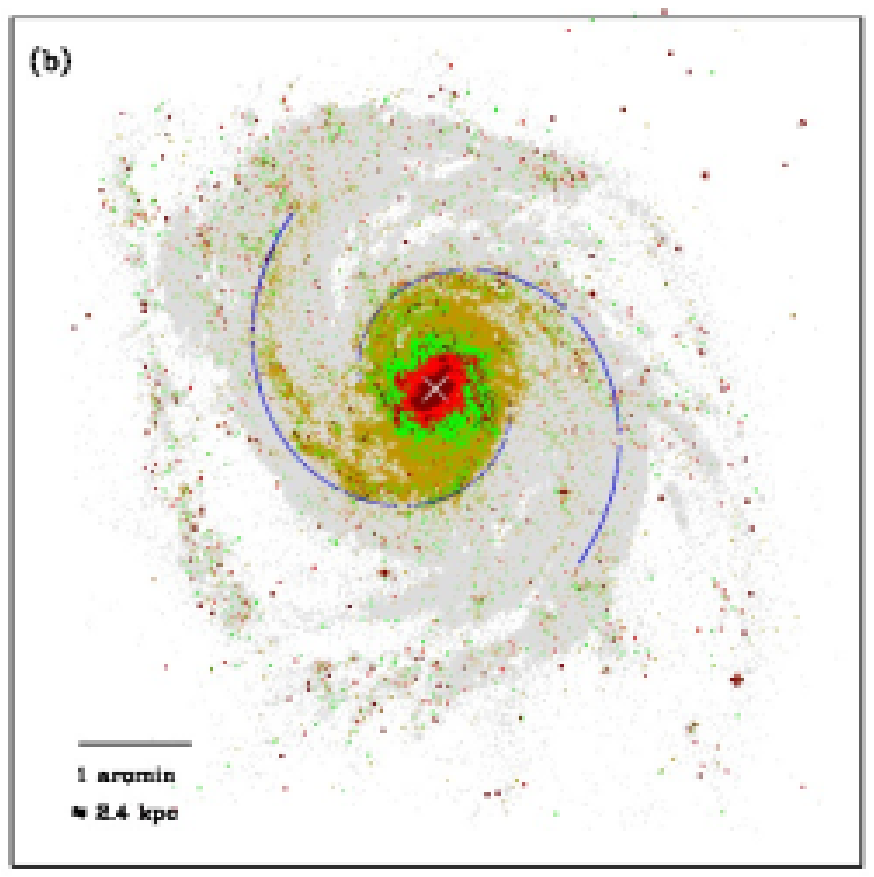}
\caption{ (a) The definition of `plumes' in the $(V-I)$ pCMD. The direction to which the plumes stretch is consistent with that of dust extinction, denoted as an arrow. The solid lines are parallel to the dust extinction arrow and their interval is 0.8, which is manually selected.
(b) The spatial distribution of plume pixels, denoted using different colors according to the domains defined in (a). The solid curves show the spiral arm patterns defined in \S\ref{armdef} and the cross locates the center of NGC 5194.
\label{plume}}
\end{figure}

It is noted that there are some `plume' features that run from the red pixel sequence to the redder domain in Fig.~\ref{pcmd}, which is particularly clear in the $V-I$ pCMD. Actually, the pixel number fraction corresponding to those plume features is relatively small compared to all the pixels in the red pixel sequence, but the plume features stretch out to very red colors (even more than 5 times the sample inter-quartile range (SIQR) at $20<V<21$ mag arcsec$^{-2}$).
In Fig.~\ref{plume}(a), the direction to which those plumes stretch seems to agree with the direction of dust extinction, which indicates that the plume features are closely related to dust extinction. Fig.~\ref{plume}(b) shows that different plumes in the pCMD correspond to different structural components. For example, the brightest plume corresponds to the central part of the bulge, while the faintest plume corresponds to the inner part of the spiral arms. One possibility for why the pixels appear to be clumped into plumes rather than smoothly distributed is that the dust distribution in and around the NGC 5194 center may be clumped and not smoothly distributed. This needs to be checked using some high resolution infrared observation, which is out of the scope of this paper.

\subsubsection{Resolution dependence}

\begin{figure*}[!t]
\plotone{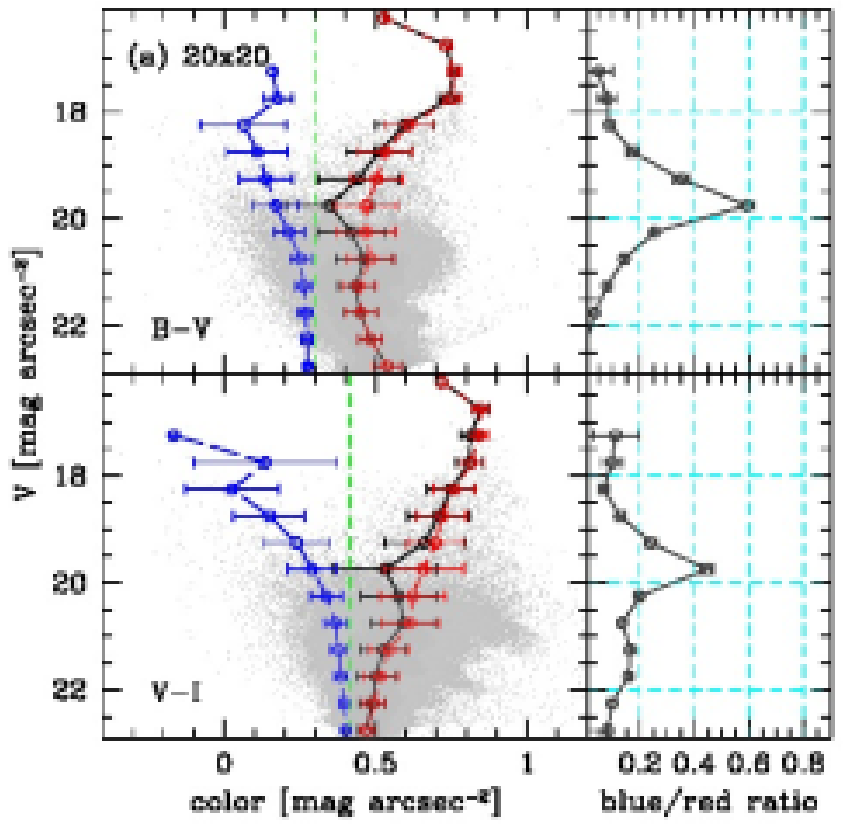}
\plotone{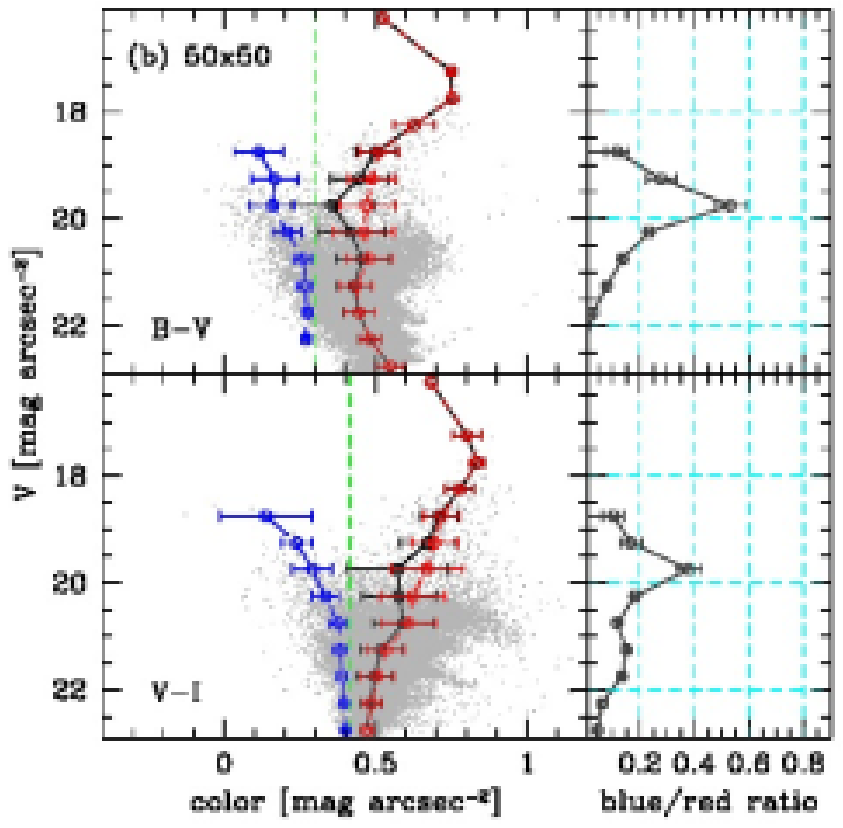}
\plotone{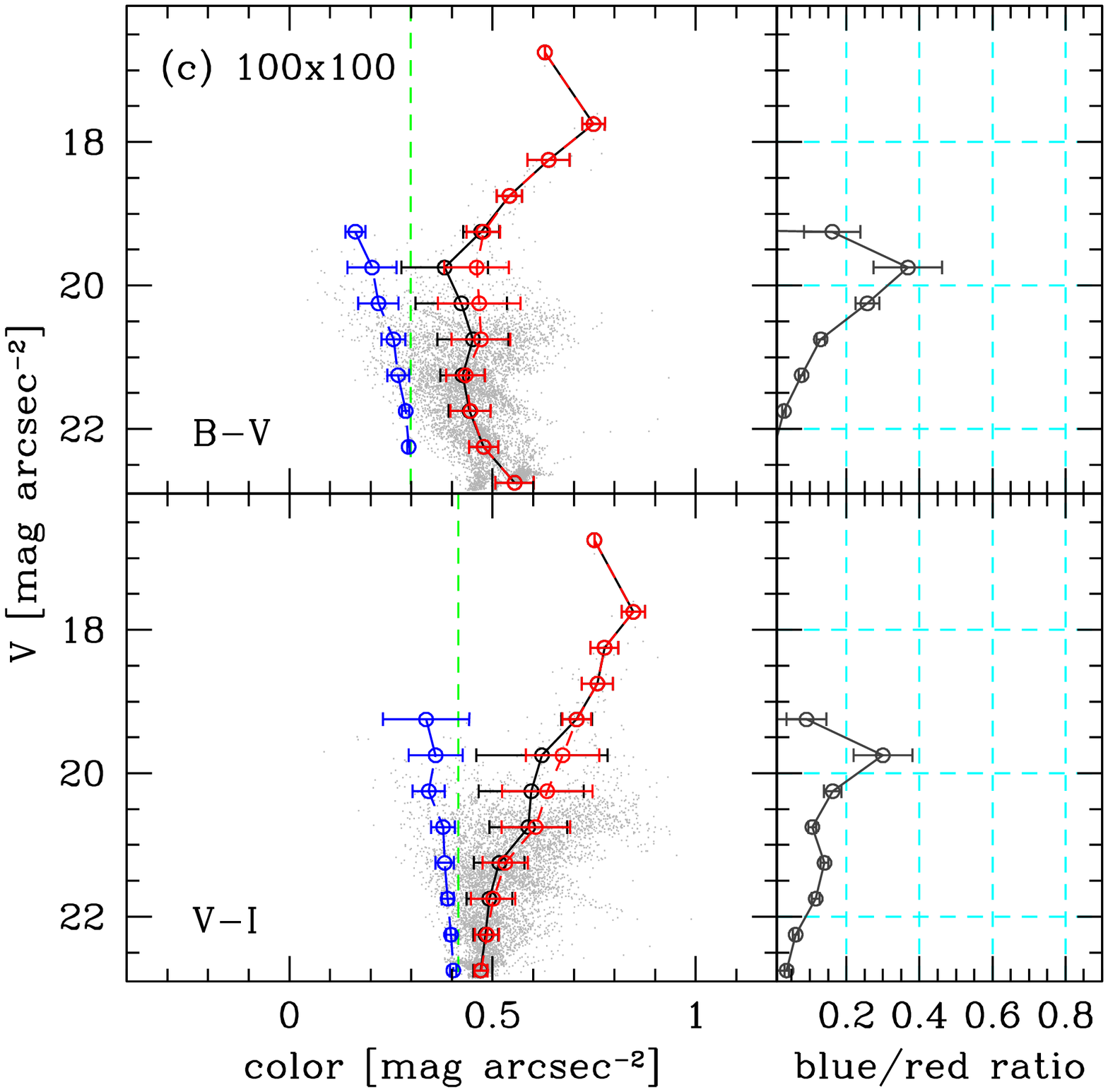}
\plotone{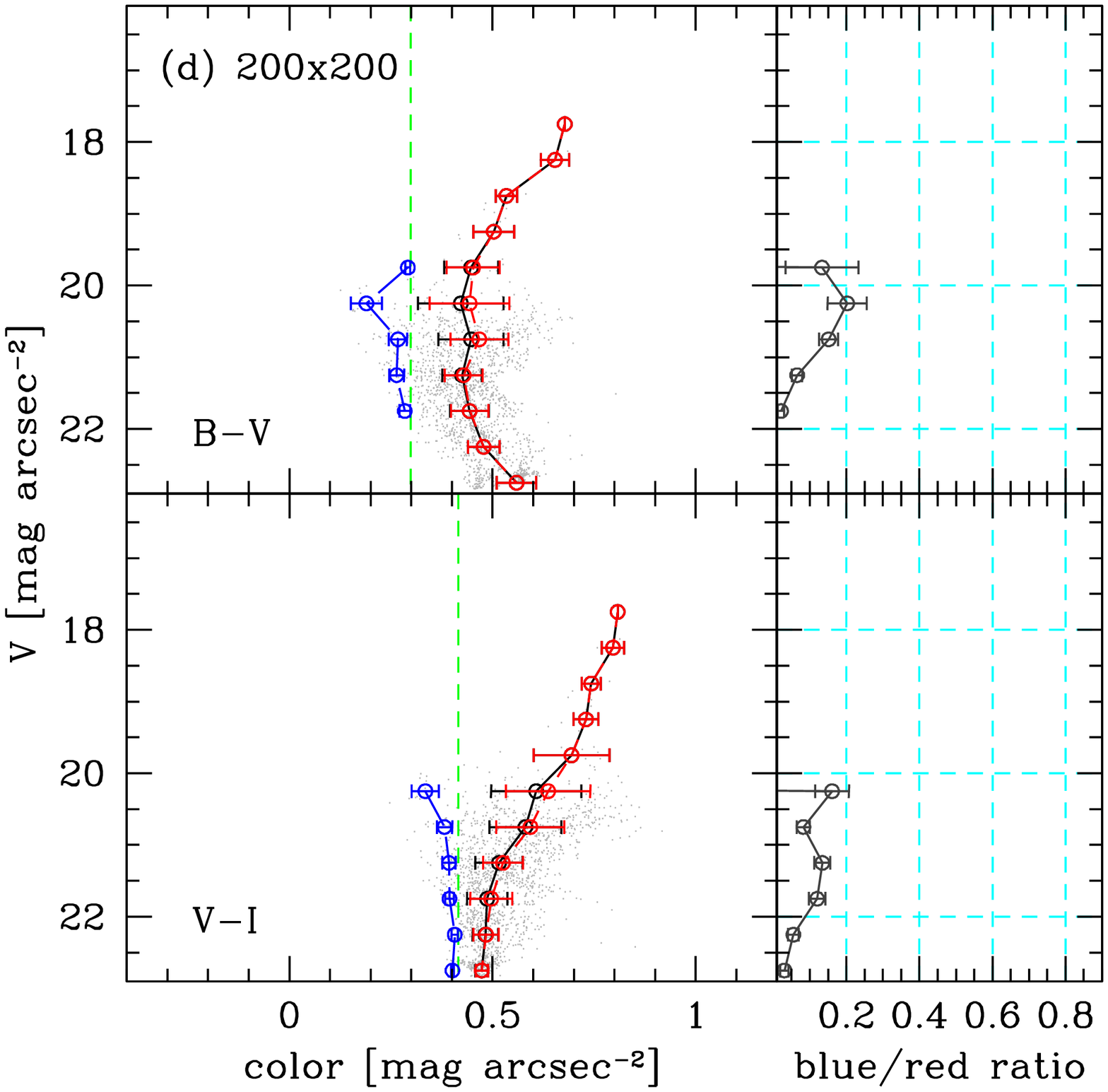}
\caption{ The same as Fig.~\ref{pcmd}, but for different binning factors: (a) $20\times20$, (b) $50\times50$, (c) $100\times100$ and (d) $200\times200$ pixels.
\label{pcmdbin}}
\end{figure*}

In \S\ref{cdist}, we derived several quantities describing the shapes of the NGC 5194 pCMDs. However, those quantities need to be compared to other galaxies for the statistical interpretation of the pCMDs related to the physical properties of galaxies\footnote{Such comparisons will be done in our future studies.}. For example, by comparing the pCMD features between different galaxies with infrared (IR) information, we can investigate how the optical pCMD features depend on the IR properties of galaxies, that is, how dust affects the shape of a pCMD. If there is a systematic pattern in the distortion of the pCMD by dust (e.g., the slope variation of the red pixel sequence), it can be inversely used as an optical indicator of dust content for galaxies without IR information.
In those comparisons, sample galaxies may have different spatial resolutions in their image data, and thus it is necessary to estimate how the pCMD features depend on spatial resolution.

Fig.~\ref{pcmdbin} shows the variation of the NGC 5194 pCMDs with different binning factors: $20\times20$, $50\times50$, $100\times100$ and $200\times200$. Since the linear scale of a pixel with $4\times4$ binning is approximately 8 pc pixel$^{-1}$, those binning factors correspond to 40, 100, 200 and 400 pc pixel$^{-1}$ resolutions, respectively. The $20\times20$ binning (40 pc pixel$^{-1}$) corresponds to the best spatial resolution using a ground-based telescope ($0.5''$) for NGC 5194 or the typical spatial resolution of the {\it HST} ($0.1''$) for an object at z = 0.02.
When we observe an object at z = 0.02 using ground-based telescopes with the best seeing condition ($0.5''$), its pCMD will be like the $100\times100$ binning result in Fig.~\ref{pcmdbin}.

The color dispersion in the pCMD decreases as the spatial resolution becomes poor, but the blue and red pixel sequences and the blue/red ratio peak are distinguishable clearly to $50\times50$ binning and marginally to $100\times100$ binning. In the pCMD with $200\times200$ binning, however, those features are smoothed out and hardly distinguishable. Thus, if we want to see the blue/red pixel sequences clearly, we need at least 100 parsec resolution.
Tables~\ref{pcmdvar1} -- \ref{pcmdvar3} list the blue and red pixel sequence parameters and the blue/red ratio as a function of the spatial resolution.

\begin{figure}[!t]
\plotone{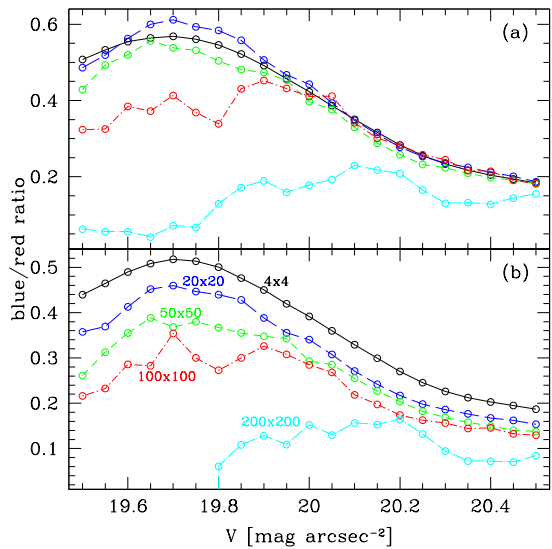}
\caption{ Blue/red ratio peak variation along different binning factors using (a) $B-V$ color cut and (b) $V-I$ color cut: $4\times4$ (solid line), $20\times20$ (long-dashed line), $50\times50$ (short-dashed line), $100\times100$ (dot-short-dashed line), $200\times200$ (dot-long-dashed line).
\label{peakcomp}}
\end{figure}

Fig.~\ref{peakcomp} shows the blue/red ratios as a function of pixel surface brightness to estimate the exact pixel surface brightness where the peak blue/red ratio is found ($\mu_{peak}$) for various resolutions. The blue/red ratio was calculated with pixels in the 0.5 mag arcsec$^{-2}$ surface brightness interval, in which the central surface brightness varies in a step of 0.05 mag arcsec$^{-2}$.
For $4\times4$ binning, the $\mu_{peak}$ is commonly 19.7 mag arcsec$^{-2}$ both when using the $B-V$ color cut and when using the $V-I$ color cut. The $\mu_{peak}$ value is almost constant up to the $50\times50$ binning, showing that the $\mu_{peak}$ does not significantly depend on spatial resolution when the spatial resolution is finer than some critical value ($\sim$ 100 pc pixel$^{-1}$).
Why is this scale critical for the pixel analysis?
One possible answer is that 100 pc is the critical scale discriminating between stellar-scale structures and galaxy-scale structures. The sizes of typical star clusters are smaller than 10 pc and even extraordinarily large star clusters like faint fuzzy clusters in M51 are smaller than 15 pc \citep{hwa06}. 100 pc is a scale larger than the typical star cluster scale by the order of 1, from which multiple stellar components in some galaxy-scale structures may start to be singled out significantly.
Thus, spatial resolution sparser than 100 pc pixel$^{-1}$ is not appropriate for pixel analysis.

\subsection{Spatial Variation of Stellar Populations}\label{svsp}

\subsubsection{Stellar population division based on the pCCD}\label{pixdiv}

Since the linear scale of a $4\times4$ binned pixel is 8 parsecs, the light of many stars may be integrated in a binned pixel. Basically, the stars in each pixel may not be a simple stellar population (SSP), but a mixture of old stars and young stars at almost every pixel. However, it is fundamentally impossible to estimate the exact star formation history only using 3-band colors: only if we suppose 2 SSPs, we need to determine at least 6 quantities (age and metallicity of population 1, age and metallicity of population 2, the ratio between populations 1 and 2, and dust extinction) from two colors. In this case, a more practical approach is to estimate the SSP-equivalent values, which represent the \emph{luminosity-weighted mean values} \citep[e.g.][]{smi09,du10,pri11}.

If we suppose that the stars in a given pixel are an SSP (that is, if they have the same age and metal abundance), we can estimate their approximate age and metallicity by comparing their colors with population synthesis models.
However, since the effects of age and metallicity on the optical colors degenerate, such an estimation still needs some assumptions, like supposing that the stars have solar metallicity. In addition, the optical colors are significantly affected by dust attenuation. Thus, in principle, it is very difficult to estimate the SSP-equivalent age and metallicity and dust attenuation of the individual pixels using only two optical colors.

\begin{figure}[!t]
\plotone{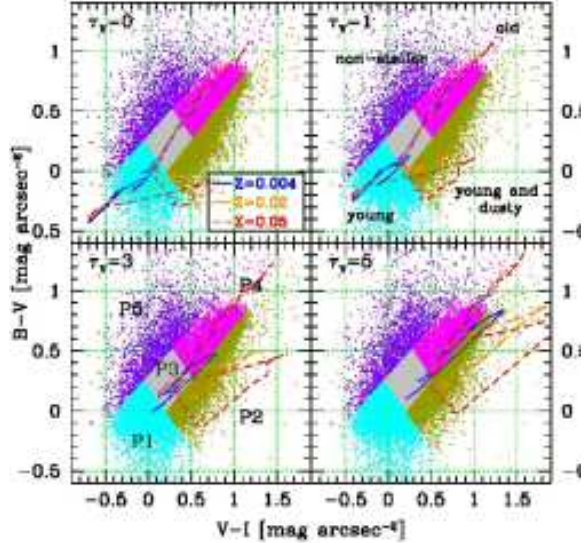}
\caption{ Pixel color-color diagrams (pCCDs) with simple stellar population (SSP) models \citep{bru03}. Dots are the pixels with $V<20$~mag~arcsec$^{-2}$ and the lines are SSP models with different metal abundances (Z = 0.02 is the solar abundance). Each panel shows the models with a different total effective optical depth in the $V$ band ($\tau_V$) and the fixed fraction of the ambient interstellar medium contribution ($\mu=0.3$), based on the simple two-component model of \citet{cha00}. The age of the SSP model at the lower-left end in each panel is about 0.1 Myr and that at the upper-right end is about 10 Gyr. The pixels are approximately divided into 5 populations: P1 (young), P2 (young and dusty), P3 (intermediate), P4 (old) and P5 (non-stellar).
\label{pccd}}
\end{figure}

In spite of such difficulties, however, it is still possible to statistically estimate those quantities of a large number of pixels with some reasonable assumptions. Fig.~\ref{pccd} shows the ($4\times4$ binned) pixel color-color relation in NGC 5194 with several population synthesis models \citep{bru03}.
Although all the factors, age, metallicity and dust attenuation, affect the optical colors, the pCCD is approximately divided into several domains for different stellar populations. For example, the domain with blue $B-V$ and $V-I$ colors (P1) is for very young stellar populations regardless of their metallicity and not for populations with strong dust attenuation ($\tau_V\ge3$). The domain with blue $B-V$ color but red $V-I$ color (P2) is for young and dusty stellar populations and the domain with red $B-V$ and $V-I$ colors is for old stellar populations for moderate metallicity and dust attenuation (P4). The domain with blue $V-I$ but red $B-V$ color (P5) seems to be for non-stellar pixels, of which the fraction is small. This pixel division is not rigorous, but useful for the statistical investigation of stellar population distribution. The population division criteria used in this paper are as follows:
\begin{eqnarray}
\lefteqn{\textrm{P1}: B-V < 0.8\times(V-I)+0.3}
\nonumber\\
& & \textrm{and } B-V < -(V-I)+0.2,\\
\lefteqn{\textrm{P2}: B-V < 0.9\times(V-I)-0.2}
\nonumber\\
& & \textrm{and } B-V \ge -(V-I)+0.2,\\
\lefteqn{\textrm{P3}: B-V \ge 0.9\times(V-I)-0.2}
\nonumber\\
& & \textrm{and } B-V < 0.8\times(V-I)+0.3
\nonumber\\
& & \textrm{and } B-V \ge -(V-I)+0.2
\nonumber\\
& & \textrm{and } B-V < -(V-I)+0.8,\\
\lefteqn{\textrm{P4}: B-V \ge 0.9\times(V-I)-0.2}
\nonumber\\
& & \textrm{and } B-V < 0.8\times(V-I)+0.3
\nonumber\\
& & \textrm{and } B-V \ge -(V-I)+0.8,\\
\lefteqn{\textrm{P5}: B-V \ge 0.8\times(V-I)+0.3.}
\end{eqnarray}

In most conditions (i.e., almost regardless of metallicity and dust), P1 pixels represent stellar populations younger than 100 Myr, although young ($<$ 100 Myr) populations do not always belong to P1. Meanwhile, P2 pixels represent stellar populations younger than 10 Myr with significant dust content ($\tau_V\gtrsim1$), although young and dusty but very metal poor (e.g., Z = 0.0001) populations do not belong to P2.
On the other hand, the P3 (intermediate-domain) pixels may be metal-rich populations with age of 100 Myr -- 1 Gyr and $\tau_V=0$ or metal-poor populations with ages of 100 Myr -- 1 Gyr and $\tau_V=5$ or metal-poor populations with ages of 1 Gyr -- 10 Gyr and $\tau_V=0$, which shows extremely strong degeneracy of age, metallicity and dust extinction. Similarly, the P4 pixels may be metal-rich populations with ages of 1 Gyr -- 10 Gyr and $\tau_V=0$ or metal-poor populations with ages of 1 Gyr -- 10 Gyr and $\tau_V=5$ or metal-rich populations with ages of 100 Myr -- 1 Gyr and $\tau_V=5$. Thus, the P4 pixels have a high probability of being old populations, but they are not necessarily old.
This is the limit of the two-color-based classification, and the pixel classification in Fig.~\ref{pccd} is just an approximation. However, the P1 and P2 pixels are relatively good approximations of a young stellar population and young and a dusty stellar population, respectively.
Thus, our analysis on the population variation mainly focuses on these two pixel populations.

\begin{figure}[!t]
\plotone{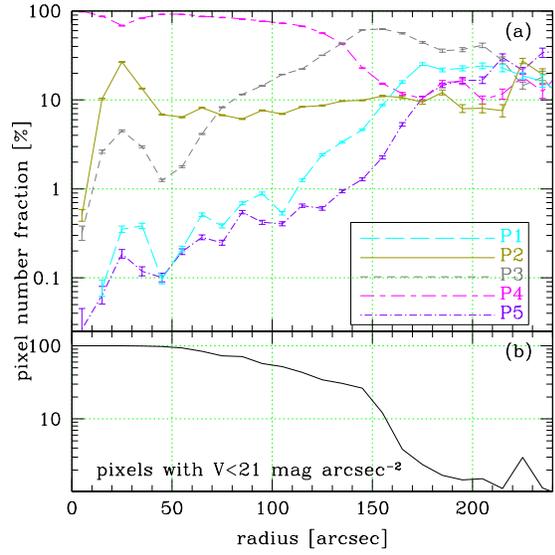}
\caption{ (a) The number fractions of the five populations in NGC 5194 as a function of radius, for pixels with $V<21$ mag arcsec$^{-2}$. The errorbars show the Poisson errors. (b) The number fraction of the pixels with $V<21$ mag arcsec$^{-2}$ to the entire pixels, as a function of the radius.
\label{radfrac}}
\end{figure}

\begin{figure}[!t]
\plotone{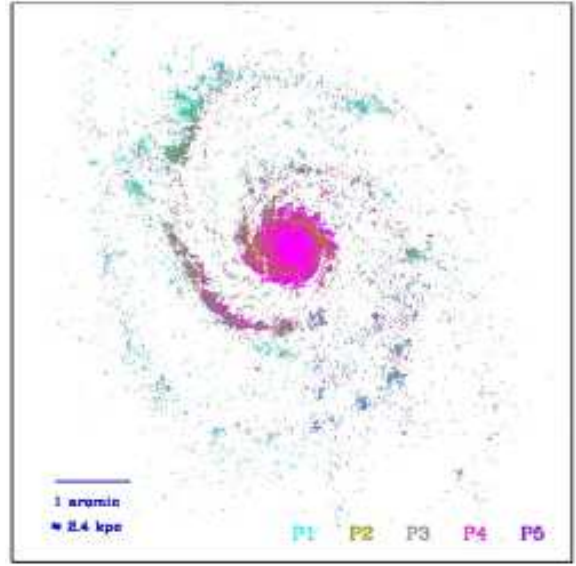}
\caption{ Spatial distribution of the five populations in NGC 5194, for pixels with $V<20$ mag arcsec$^{-2}$.
\label{popmap}}
\end{figure}

\begin{figure}[!t]
\plotone{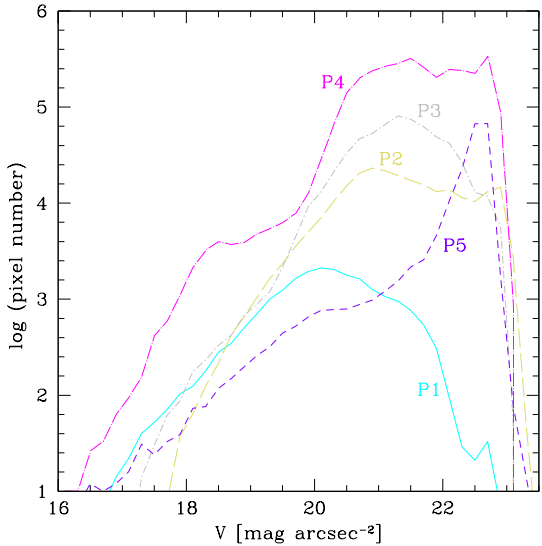}
\caption{ $V$ band surface brightness distribution of each pixel population: P1 (solid), P2 (long-dashed), P3 (dot-short-dashed), P4 (dot-long-dashed) and P5 (short-dashed).
\label{pixelcount}}
\end{figure}

Fig.~\ref{radfrac}(a) shows the number fraction of each population defined in Fig.~\ref{pccd} as a function of NGC 5194's radius.
The old population (P4) is mainly distributed at the inner area of NGC 5194 within $120''$ ($\approx$ 4.8 kpc), while the young population (P1) increases with the radius until $\sim180''$ ($\approx$ 7.2 kpc). It is interesting that the young and dusty population (P2) is almost constantly distributed at radius $>40''$ ($\approx$ 1.6 kpc) but it shows a peak at $\sim25''$ ($\approx$ 1 kpc).
The peak of P2 is due to the existence of the dusty inner ring of NGC 5194 \citep{pie86} as shown in Fig.~\ref{popmap}, which seems to be connected to the dusty populations in the spiral arms.

The P5 fraction is very small ($<0.1\%$) in the NGC 5194 center, but it increases with the radius, particularly rapidly at R $\gtrsim150''$ (from $2\%$ at R $\sim150''$ to $>20\%$ at R $>200''$), although the number fraction of pixels with $V<21$ mag arcsec$^{-2}$ is very small at such a large radius, as shown in Fig.~\ref{radfrac}(b). This result gives a hint about the origin of the P5 color, implying that the P5 may be related to background light. For example, some background galaxies at high redshifts have colors like those of the P5 pixels \citep[i.e., blue $V-I$ color but red $B-V$ color; e.g.][]{hil05}. The color of the P5 pixels, which is not easily explained using population synthesis models at z = 0, may be the result of the mixture of the faint extended light from NGC 5194 and the background light from high-redshift galaxies.
Fig.~\ref{pixelcount} displays the surface brightness distribution of each pixel population, showing that most of the P5 pixels are significantly biased to faint surface brightness ($V\gtrsim22$ mag arcsec$^{-2}$). The results in Fig.~\ref{radfrac}(b) and Fig.~\ref{pixelcount} confirm that the P5 pixels are biased to the outskirt and faint pixels, which supports the possibility that the P5 pixels are contaminated by background light.

The other (P1 -- P4) pixels may also be partially contaminated by background light. It is very difficult to quantify by how much, but we discuss it qualitatively here.
The color vector of such contaminations may be to the direction of bluer $V-I$ and redder $B-V$ colors (from lower-right to upper-left in Fig.~\ref{pccd}), which is almost independent of the division between P1, P3 and P4 pixels. On the other hand, the P2 pixels may appear to be P3 or P4 pixels by the contamination. However, such an effect (1) will be very weak for bright pixels (e.g., $V<21$ mag arcsec$^{-2}$; the pixel surface brightness limit in \S\ref{popvartext}) and (2) will make the population variations be seen weaker rather than stronger.
In short, the effect of the background light on our results may be very limited and our results may show the lower limits of the population variations.

\subsubsection{NGC 5194 area division}\label{armdef}

\begin{figure}[!t]
\plotone{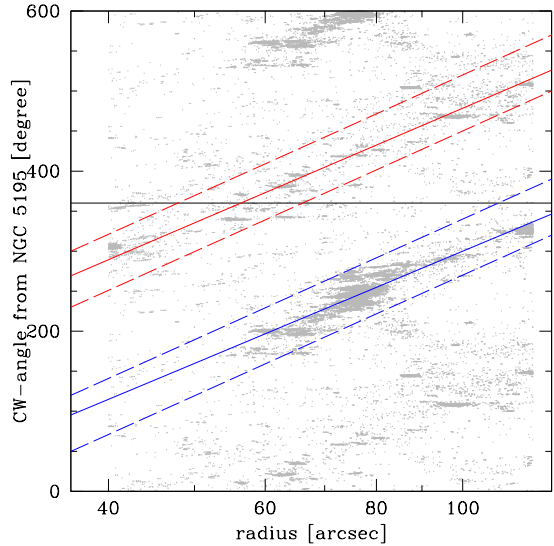}
\caption{ Spiral arms definition in the clockwise(CW)-angle from NGC 5195 versus radius plot for pixels with $V<20$ mag arcsec$^{-2}$. The radius is in logarithmic scale. The dashed lines are manually-selected limits of the two spiral arms in NGC 5194, within which the spiral arms (solid lines) are defined using the linear least-squares fit method. The horizontal solid line indicates $360^{\circ}$, the NGC 5195 direction.
\label{spiral}}
\end{figure}

\begin{figure}[!t]
\plotone{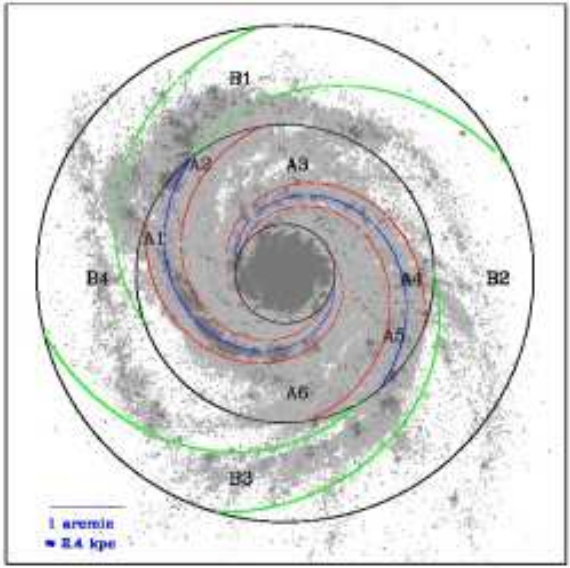}
\caption{ Divided areas for the comparison of populations. The three concentric circles are centered on the NGC 5194 center and have radii of 40, 120 and 200 arcseconds, respectively. {\it Inner ring}: The spiral arms were defined in Fig.~\ref{spiral} and the $\pm30^{\circ}$ from the spiral arms are set as the boundary, dividing six areas: A1 -- A6. {\it Outer ring}: Two spiral arms are neither perfectly symmetric nor perfectly spiral. Thus, the area boundary is set manually so that `B1 -- B2' and `B3 -- B4' are approximately symmetric in the spiral structures.
\label{areadiv}}
\end{figure}

Here, we focus on the population variations in two viewpoints: 1) the variation across the spiral arms and 2) the difference between the NGC 5195 direction and the opposite direction. To investigate these variations, the spiral arms of NGC 5194 need to be defined reasonably. We defined the spiral arms by fitting the spatial distribution of bright ($V<20$ mag arcsec$^{-2}$) pixels. A spiral pattern is linearly distributed in the angle versus log radius plot, but the spiral arms of NGC 5194 are significantly distorted at the outskirt area due to the interaction with NGC 5195. Thus, we fit the spiral arms only in the radius range of 40 -- 120 arcseconds (1.6 -- 4.8 kpc) as shown in Fig.~\ref{spiral}, which returns the spiral arm coordinate equations of:
\begin{equation}\label{spa1}
\Theta = 465.056 \times \log R - 630.692
\end{equation}
and
\begin{equation}\label{spa2}
\Theta = 475.812 \times \log R - 473.584,
\end{equation}
where $\Theta$ is the clockwise position angle from the NGC 5195 direction in the unit of degree and $R$ is the distance to the NGC 5194 center in the unit of arcsecond. Equation~\ref{spa1} is for the spiral arm between A1 and A2 (hereafter SA1) and Equation~\ref{spa2} is for that between A4 and A5 (SA2) in Fig.~\ref{areadiv}.

Based on the defined spiral arm location, we selected ten areas as shown in Fig.~\ref{areadiv}. At $1.6<R<4.8$ kpc, the A1 - A2 and A4 - A5 cover the $\pm30^{\circ}$ areas of SA1 and SA2, respectively, and A3 and A6 areas are the inter-arm areas. Supposing that the spiral arms of NGC 5194 are trailing arms and considering that the rotation velocity of spiral patterns are known to be slower than inner disk stars, we refer to A1 and A4 as the areas `after' the spiral arms and A2 and A5 as the areas `before' the spiral arms, in the sense that the relative motion of stars to the spiral arms is along A2 $\rightarrow$ A1 (or A5 $\rightarrow$ A4).

At $4.8<R<8$ kpc, the B1 - B2 and B3 - B4 areas are the outskirt spiral arms, distorted by the tidal interaction with NGC 5195.
These areas were set manually, referring to the spiral arm equations (\ref{spa1} and \ref{spa2}). The equations for the boundaries between the areas are as follows:
\begin{eqnarray}
\textrm{B1/B2}: \Theta = 465.056 \times \log R - 630.692,\\
\textrm{B2/B3}: \Theta = 475.812 \times \log R - 523.584,\\
\textrm{B2/B3}: \Theta = 475.812 \times \log R - 463.584,\\
\textrm{B4/B1}: \Theta = 465.056 \times \log R - 700.692.
\end{eqnarray}

\subsubsection{Population Variation}\label{popvartext}

\begin{figure}[!t]
\plotone{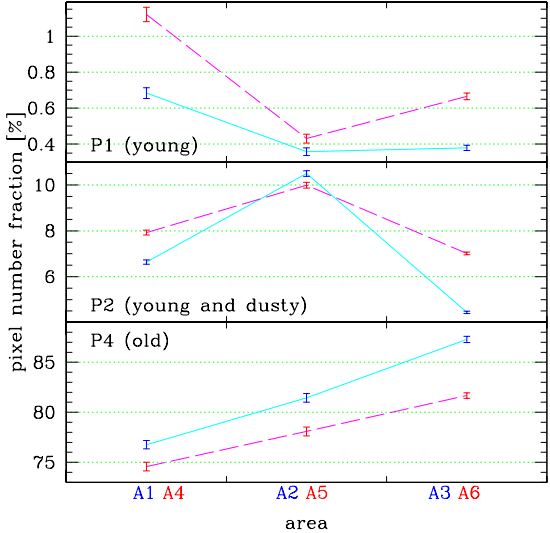}
\caption{ Population variations at the radius of 40 -- 120 arcseconds (1.6 -- 4.8 kpc), for P1 (young), P2 (young and dusty) and P4 (old) populations with $V<21$ mag arcsec$^{-2}$. The solid lines show the variations along the A1 -- A2 -- A3 areas and the dashed lines show those along the A4 -- A5 -- A6 areas. The errorbars indicate the Poisson errors.
\label{iring}}
\end{figure}

Fig.~\ref{iring} shows the population variation across the spiral arms. The fractions of the populations are slightly different between the two spiral arms, but their variation trends are strikingly consistent. Before and after the spiral arms, the fraction of the young population (P1) increases twice or more ($0.35\%$ in A2 $\rightarrow$ $0.68\%$ in A1; $0.43\%$ in A5 $\rightarrow$ $1.12\%$ in A4). The difference between the before-arm areas and inter-arm areas are relatively small.
On the other hand, the young and dusty population (P2) is most dominant in the before-arm areas ($\sim10\%$ in A2 and A5, but $4-8\%$ in other areas). The old population (P4) shows an increasing fraction along after-arm $\rightarrow$ before-arm $\rightarrow$ inter-arm areas.

\begin{figure}[!t]
\plotone{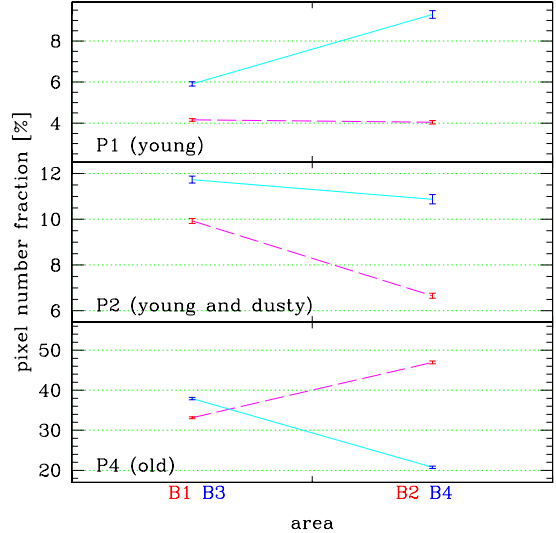}
\caption{ Population variations at the radius of 120 -- 200 arcseconds (4.8 -- 8.0 kpc), for P1 (young), P2 (young and dusty) and P4 (old) populations with $V<21$ mag arcsec$^{-2}$. The dashed lines show the variations along the B1 -- B2 areas and the solid lines show those along the B3 -- B4 areas. The errorbars indicate the Poisson errors.
\label{oring}}
\end{figure}

Fig.~\ref{oring} compares the outskirt populations between the two spiral arms. In the outskirt area of SA1 (B1 and B2), the fraction of the young population (P1) is almost constant ($\sim4\%$), whereas that in B4 ($9.3\%$)is significantly larger than that in B3 ($5.9\%$; by more than 1.5 factor). We also found that the fraction of the old population (P4) shows opposite trends between the outskirts of SA1 and SA2: the P4 fraction along B3 $\rightarrow$ B4 decreases almost by half ($38\%\rightarrow21\%$), while that along B1 $\rightarrow$ B2 increases by a 1.4 factor ($33\%\rightarrow47\%$).
The fraction of the young and dusty population (P2) decreases along the outward direction of spiral arms both in SA1 and SA2, but the decrease along B3 $\rightarrow$ B4 ($11.7\%\rightarrow10.9\%$) is smaller than that along B1 $\rightarrow$ B2 ($9.9\%\rightarrow6.7\%$). Since the B4 area is connected to NGC 5195, the interaction with NGC 5195 seems to have affected the stellar populations in B4.

\subsection{Bright-End Pixels}\label{bep}

\begin{figure}
\plotone{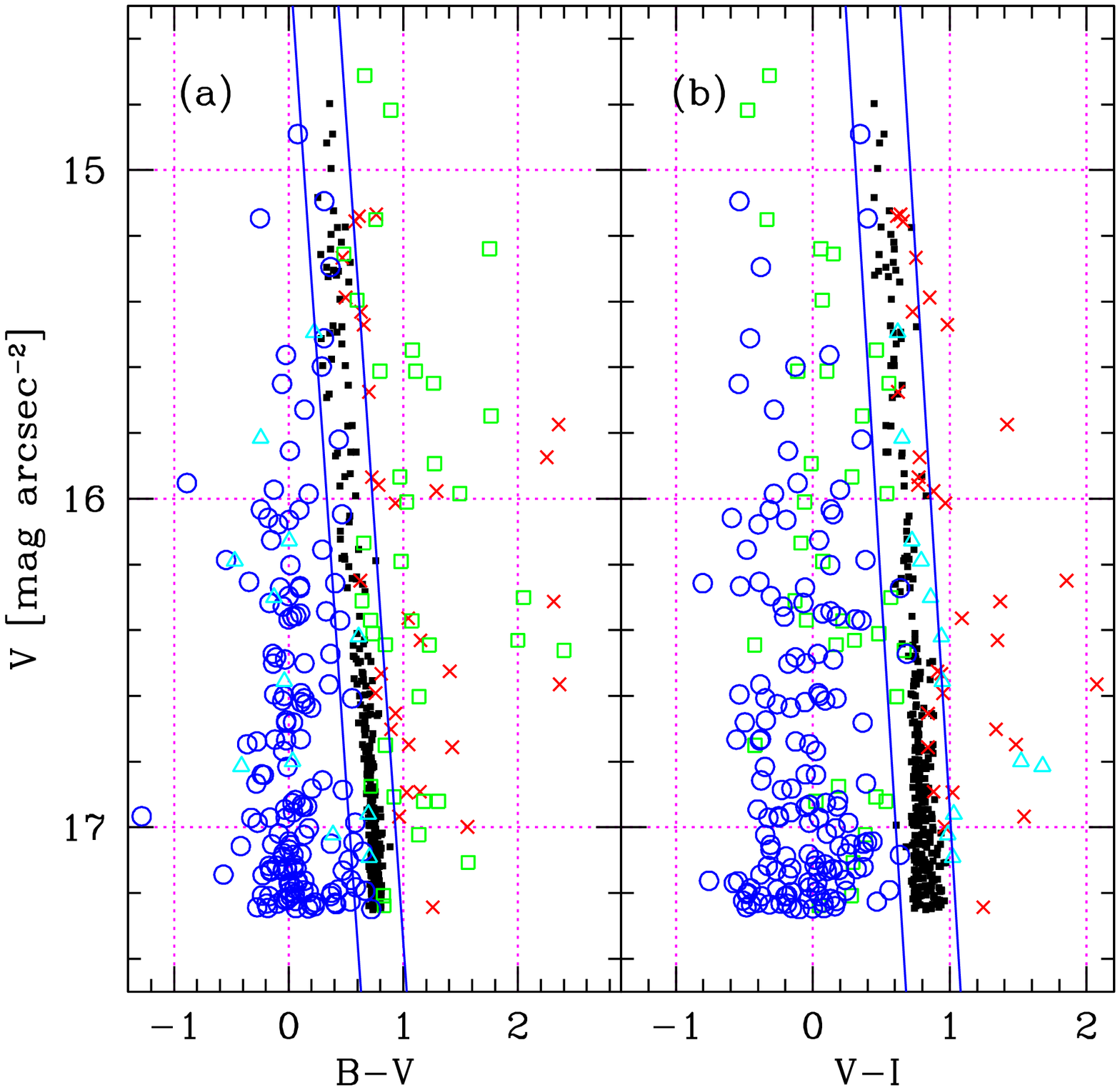}
\plotone{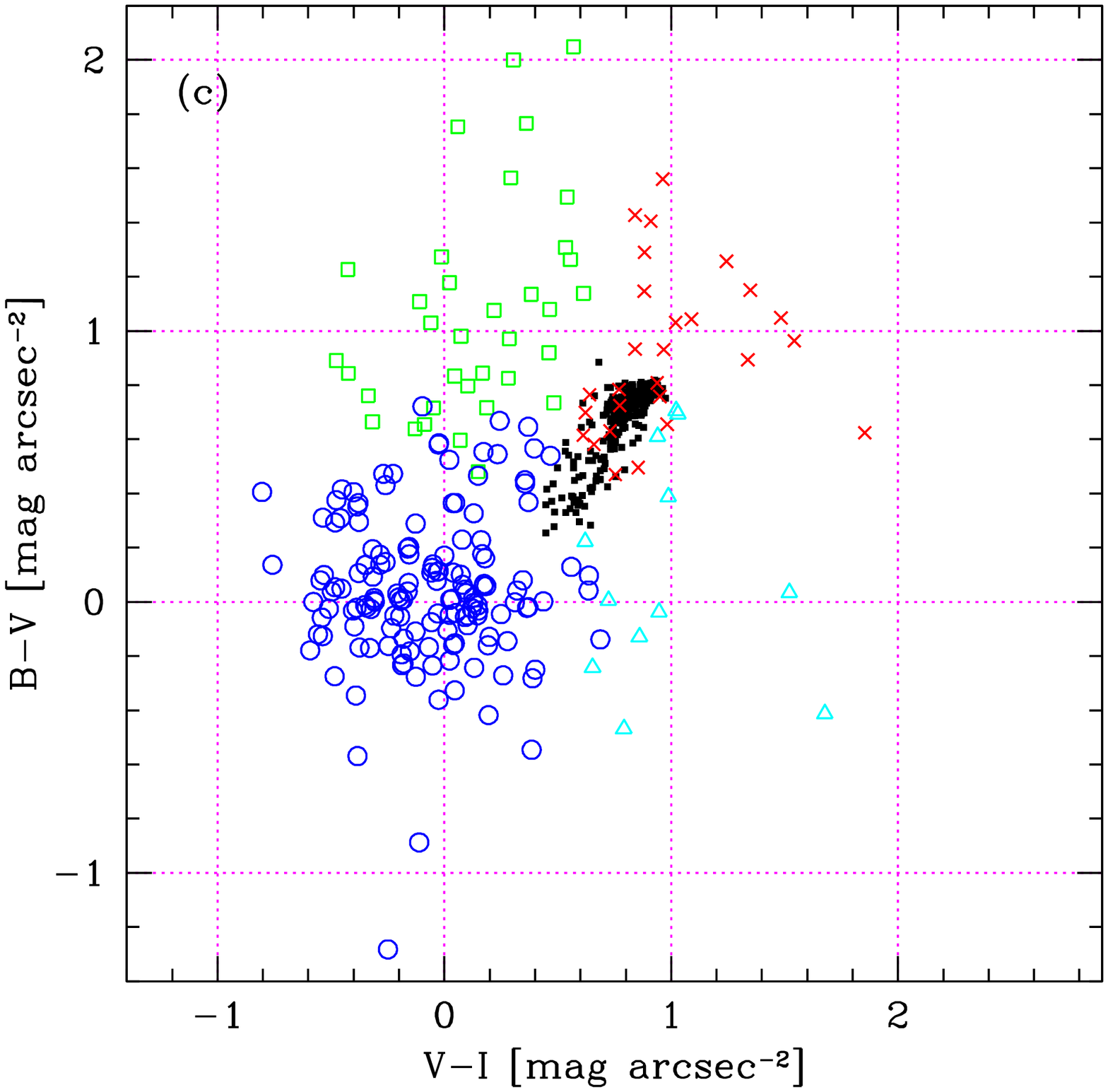}
\caption{ The bright-end ($V<17.25$ mag arcsec$^{-2}$) pixels of NGC 5194: (a) $V$ versus $B-V$, (b) $V$ versus $V-I$ and (c) $B-V$ versus $V-I$ diagrams. The primary sequence is fit by the linear least-squares fitting method and the bright-end sequence population (BSP; black dots) is defined as the pixels within a narrow cylinder in the $V$-$(B-V)$-$(V-I)$ domain, with a 0.2 radius of the color indices (solid lines). The non-BSPs are divided into four populations according to their color range: redder than BSP in both $B-V$ and $V-I$ (red crosses); bluer than BSP in both $B-V$ and $V-I$ (blue circles); redder in $B-V$ but bluer in $V-I$ (green rectangles); and bluer in $B-V$ but redder in $V-I$ (cyan triangles).
\label{btpop}}
\end{figure}

\begin{figure}[!t]
\plotone{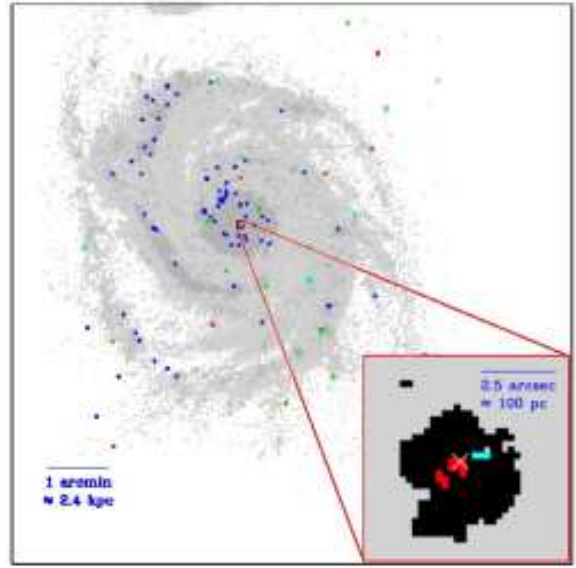}
\caption{ Spatial distribution of the bright-end ($V<17.25$ mag arcsec$^{-2}$) pixels of NGC 5194. The dot colors are as defined in Fig.~\ref{btpop}, except for the gray dots ($V<17.25$ mag arcsec$^{-2}$). The center of NGC 5194 is marked as a white cross (RA = 13h 29m 52.72s, Dec = +47d 11m 43.4s, J2000).
\label{btmap}}
\end{figure}

\begin{figure}[!t]
\plotone{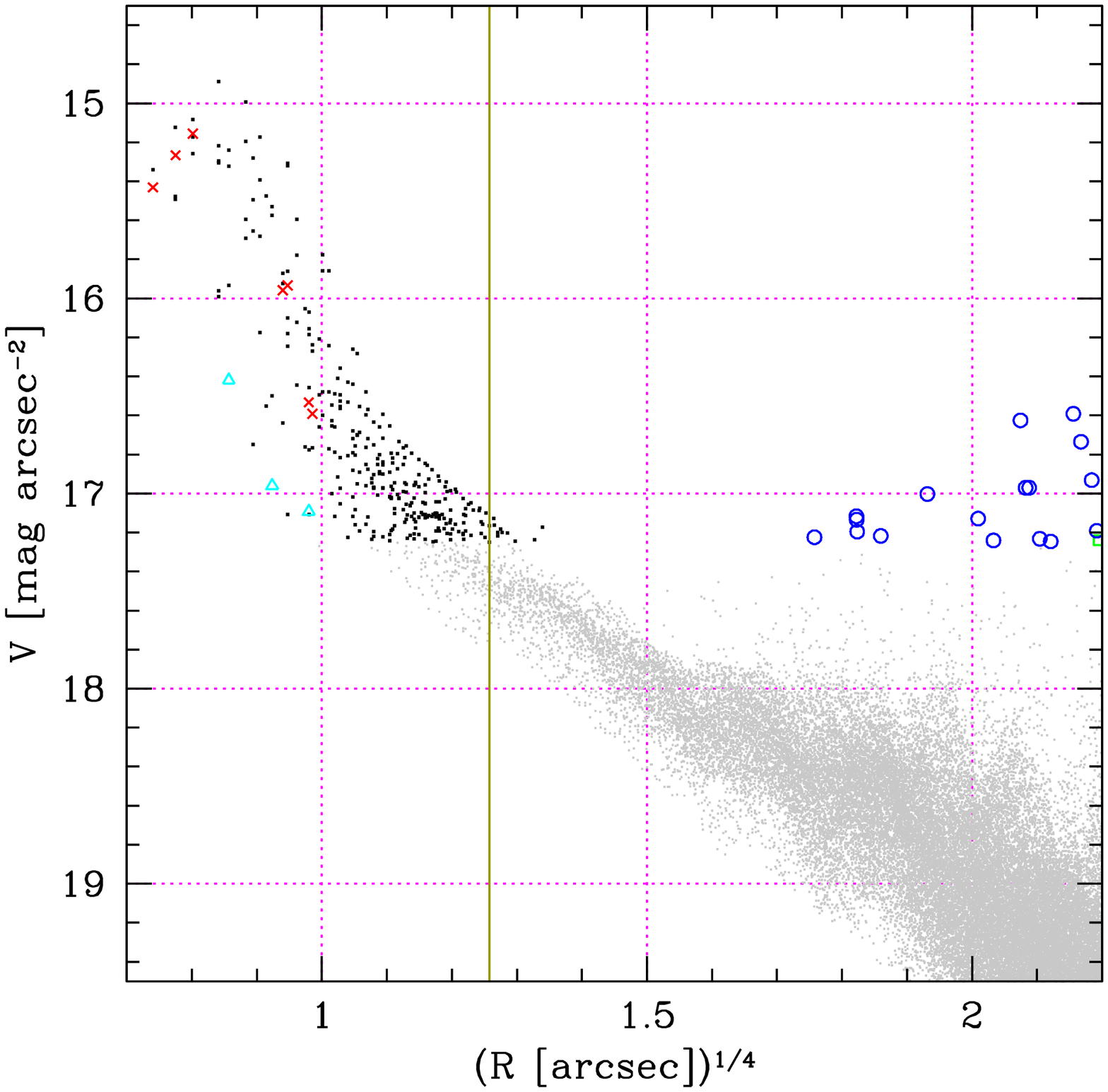}
\caption{ Pixel surface brightness versus R$^{\frac{1}{4}}$ in the central region of NGC 5194. The dot symbols and colors are defined in Fig.~\ref{btpop}, except for the gray dots ($V<17.25$ mag arcsec$^{-2}$). The vertical line is drawn at R = 2.5$''$ ($\approx100$ pc).
\label{btprof}}
\end{figure}

In the NGC 5194 pCMD, one interesting feature is found at the bright-end of the red pixel sequence: the sequence turns into the opposite direction, resulting in bluer colors for brighter pixels.
To investigate the nature of those bright pixels, we first divided the bright-end ($V<17.25$ mag arcsec$^{-2}$) pixels using their colors.
Fig.~\ref{btpop} shows the pCMDs and pCCD of the NGC 5194 bright-end pixels. There is a population forming a tight sequence in the pCMD (bright-end sequence population; BSP), which is defined as the pixels within a narrow cylinder in the 3-dimensional domain of $V$ - $(B-V)$ - $(V-I)$. The cross section of the cylinder has a radius of 0.2 in the plane of $(B-V)$ - $(V-I)$ at a given $V$, and the center of the cross section as a function of $V$ is defined as:
\begin{equation}
B-V = 0.198 \times V - 2.64
\end{equation}
\begin{equation}
V-I = 0.147 \times V - 1.70,
\end{equation}
which were derived using linear least-squares fitting.
The bright pixels out of the cylinder were divided into four populations as shown in Fig.~\ref{btpop}(c): pixels red both in $B-V$ and $V-I$ (red dots), pixels red in $B-V$ but blue in $V-I$ (green dots), pixels blue both in $B-V$ and $V-I$ (blue dots) and pixels blue in $B-V$ but red in $V-I$ (cyan dots).
Note that the pCMD slope of these bright pixels ($\Delta (V-I)/\Delta V = 0.147\pm0.006$) is different from the slope by dust extinction ($\Delta (V-I)/\Delta V = 0.226$; see Fig.~\ref{plume}), which implies that the BSP sequence slope is not explained using dust extinction only.

The spatial distribution of those bright pixels is shown in Fig.~\ref{btmap}. The pixels blue both in $B-V$ and in $V-I$ (blue dots) are distributed along the spiral arms and the dusty ring area, but the other `non-BSP' pixels are scattered irregularly. The most impressive result is that the BSP pixels are highly concentrated on the NGC 5194 center within $\sim100$ parsecs in radius. In the lower-right box of Fig.~\ref{btmap} (the zoomed-up map of the NGC 5194 center), the BSP pixels are compactly concentrated but are divided into two areas by the non-BSP pixels that seem to form a dust lane.
It is known that an active galactic nucleus (AGN) exists at the center of NGC 5194 and it was already investigated using radio and X-ray observations \citep{for85,ter01}, which may be closely related to the nature of the BSP pixels.
Fig.~\ref{btprof} shows the radial distribution of pixel surface brightness in the bulge area of NGC 5194. Whereas the pixels at R $>1''$ follow the well-known R$^\frac{1}{4}$ profile, the pixels at R $<1''$ have a very rapid slope, which implies that the central pixels may be significantly affected by the AGN.

\section{DISCUSSION}

\subsection{Interpretation of the pCMD}

Here, we discuss several parameters/features describing the pCMD of NGC 5194, which will be useful for quantitative comparisons of pCMDs between different galaxies. To establish a systematic pCMD analysis method, these parameters need to be accumulated and statistically compared between various types of galaxies in the future. The pCMD data of NGC 5194 are the beginning of such data accumulation.

\subsubsection{Blue/red color cut}

The boundary color itself between blue and red pixels is an important parameter describing the features of pCMDs. In this paper, we derived it using double Gaussian fitting of the pixels with $18.5\le V<19.5$ mag arcsec$^{-2}$, because the pixels in that surface brightness range show most clearly separated double Gaussian color distribution. It is noted that such a \emph{best} surface brightness range may not be universal for all galaxies. Actually, even for NGC 5194, different binning factors return different \emph{best} surface brightness ranges, because the pixel surface brightness tends to be squashed (i.e., bright pixels become less bright while faint pixels become brighter) when they are binned with large factors. Thus, it is best to select the \emph{best} surface brightness range case by case, which shows the most clearly separated double Gaussian distribution.
Since our blue/red color cut is based on the red Gaussian (i.e., the red pixel sequence) not the blue Gaussian, it is relatively insensitive to short-timescale activity like recent star formation. Instead, it is a parameter reflecting the metal abundance and dust content of NGC 5194, as the red pixel sequence does.

\subsubsection{Red pixel sequence}\label{disrps}

\citet{lan07} showed that red pixel sequences\footnote{\citet{lan07} named them as \emph{prime sequences}.} are found not only in early-type galaxies, but also in some late-type galaxies. The red pixel sequences in late-type galaxies are particularly clear for bulge-dominated late-type galaxies and they have color dispersions larger than those in early-type galaxies.
The pCMD shape of NGC 5194 is approximately consistent with the late-type galaxy pCMDs described by \citet{lan07}, although their details largely depend on the individual characteristics of late-type galaxies.

Since the red and blue pixel sequences are individually defined in the $B-V$ and $V-I$ colors, the pixels in those sequences have slightly different meanings according to the criterion color. The $B-V$ red sequence pixels consist of mostly P4 pixels and some P2, P3 and P5 pixels, while the $V-I$ red sequence pixels consist of mostly P2 and P4 pixels and a small fraction of P3 and P5 pixels. However, since the number of P2 and P5 pixels is much smaller than that of P4 pixels ($f_{P4}>80\%$ at R $<90''$ and $f_{P4}>50\%$ at R $<130''$; see Fig.~\ref{radfrac}), the dominant pixel population in the red pixel sequence is the P4, for both $B-V$ and $V-I$ colors. Thus, if we suppose solar metallicity (Z = 0.02) and moderate dust extinction ($\tau_V=1$), the luminosity-weighted mean stellar ages of the red sequence pixels are mostly $\gtrsim1$ Gyr. However, some pixels in the plume features (particularly in the $V-I$ pCMD) may be very young ($<10$ Myr) and dusty populations, as discussed in \S\ref{popvartext}.

There are two major factors affecting the pCMD red pixel sequences. First, since pixels closer to the center of a galaxy tend to be brighter, a red pixel sequence reflects the color gradient along the radius of a galaxy (or a galaxy bulge). It is well known that the main origin of the color gradient of bulges or elliptical galaxies is their \emph{metallicity gradient} \citep[e.g.][]{ko05,lab09,tor10}, which is an important factor determining the shape of the red pixel sequence.
When it is supposed that the red pixel sequence slope is purely determined by metallicity variation, the estimated metallicity difference between $V=20$ mag arcsec$^{-2}$ and $V=17$ mag arcsec$^{-2}$ is as large as $\Delta$[Fe/H] $\sim2$ (from Z $\approx0.0002$ to Z $\approx0.02$; with mean stellar age = 3 Gyr and $\tau_V=1$ assumptions). However, since the typical metallicity gradients of early-type galaxies (of which properties are very similar to those of bulges) are smaller than 0.6 \citep{spo10}, this estimated metallicity difference is too large.

The second factor is internal \emph{dust attenuation}. As reported by \citet{lan07}, some elliptical galaxies show hook features in their red pixel sequences, which seems to be affected by dust. Since dust makes optical colors redder, the shape of a red pixel sequence is distorted according to the distribution of dust in a galaxy.
When it is supposed that the red pixel sequence slope is purely determined by dust attenuation, the estimated optical depth difference by dust between $V=20$ mag arcsec$^{-2}$ and $V=17$ mag arcsec$^{-2}$ is as large as $\Delta\tau_V\sim4$, corresponding to $\Delta E(B-V)\sim0.2$.

However, the actual red pixel sequence in the pCCD does not exactly agree with either the sequence of the metallicity effect or that of the dust effect. Thus, the red pixel sequence is thought to be affected by both metallicity and dust variations, and in addition, the effect of age variation may not be zero.
As shown in \citet[][see their Section 3.3.1 on the \emph{red hook} features]{lan07}, the effect of dust on the red pixel sequence slope seems to be different from that of metallicity.
Thus, if pCMD data for a large enough number (more than several tens) of galaxies with independently estimated metallicity and dust information are secured, we can establish a statistical method to distinguish the effects of metallicity and dust on the red pixel sequence slope and color offset.
After that, inversely, it will be possible to estimate the approximate metallicity and dust content of galaxies only using their pCMDs.

\subsubsection{Blue pixel sequence}

As displayed in Fig.~\ref{cmap}a, the blue pixel sequence reflects the properties of the disk and spiral arms: the star-forming area with young stars.
Like the red pixel sequence, the blue pixel sequence has slightly different meanings according to the criterion color. The pixels in the $B-V$ blue pixel sequence consist of P1, P2 and P3 pixels, while the pixels in the $V-I$ blue pixel sequence consist of P1, P3 and P5 pixels. For those populations, if solar metallicity (Z = 0.02) and moderate dust extinction ($\tau_V=1$) are supposed, the luminosity-weighted mean stellar ages of the blue sequence pixels are mostly $\lesssim1$ Gyr.
When it is supposed that the color variation in the blue pixel sequence is purely due to age variation, the estimated SSP-equivalent age of the blue pixel sequence at $V<20$ mag arcsec$^{-2}$ ranges from 5 Myr (at $V\sim16$ mag arcsec$^{-2}$) to 300 Myr (at $V\sim20$ mag arcsec$^{-2}$; with solar metallicity and $\tau_V=1$ assumptions). This pixel age estimation agrees with the result of \citet{hwa10} that the star cluster formation rate in M51 increased significantly during the period of $100-250$ Myr ago.

Since the pixels along spiral arms tend to be relatively bright ($V<20-21$ mag arcsec$^{-2}$), we infer that the brightest tip of the blue pixel sequence may be brighter for a galaxy with brighter and more obvious spiral arms. However, to use the brightest tip of the blue pixel sequence as a star formation rate indicator, it should be considered that the brightest tip strongly depends on spatial resolution, as shown in Fig.~\ref{pcmdbin} and Tables~\ref{pcmdvar1}--\ref{pcmdvar2}. In addition to the brightest tip, the color dispersion at given surface brightness may be also a possible parameter reflecting the star formation history of a galaxy.

\subsubsection{Blue/red ratio}

\citet{lan07} suggested that total blue/red light ratio of a galaxy can be used as a morphology indicator for that galaxy, showing a good correlation between the total blue/red ratio and galaxy morphological type. As an extension of such an idea, we investigated the blue/red ratio as a function of pixel surface brightness in \S\ref{cdist}.
As a result, we found that the pCMD of NGC 5194 shows a remarkable feature of the blue/red ratio peak at $19.5\le V<20.0$ mag arcsec$^{-2}$, which is consistently found even in the low-resolution versions of the pCMD up to $100\times100$ binning (200 pc pixel$^{-1}$ resolution). Interestingly, the peak blue/red ratio itself in the $B-V$ pCMD remains consistent within 1-$\sigma$ uncertainty up to the $50\times50$ binning (100 pc pixel$^{-1}$ resolution) pCMD ($0.532\pm0.061$).
However, the peak blue/red ratio in the $V-I$ pCMD tends to decrease as the binning factor increases.

The pixels around the surface brightness for the blue/red ratio peak ($\mu_{peak}\sim19.7$ mag arcsec$^{-2}$) reveal the spiral-arm patterns well, indicating that the blue/red ratio in this range is a measure of the spiral arm fraction in this galaxy.
This parameter may be a morphology indicator better than the total blue/red light ratio, because it is derived in a controlled surface brightness range.
The $\mu_{peak}$ does not depend on spatial resolution, when spatial resolution is finer than $\sim100$ pc pixel$^{-1}$. However, it needs to be checked whether $\mu_{peak}$ is universal or depending on individual properties of spiral galaxies.

In principle, the total color itself of a galaxy can not be a perfect morphology indicator, because there are some unusual types of galaxies such as blue early-type galaxies and red late-type galaxies \citep[e.g.][]{lee08,lee10a,lee10b,lee10c}. However, the combination of $\mu_{peak}$ and the blue/red ratio at $\mu_{peak}$ may be useful to discriminate blue early-type galaxies from usual late-type galaxies, because blue early-type galaxies are known to be bluest at their brightest center \citep[that is, the surface brightness distributions of their blue pixels may be different from those of usual late-type galaxies;][]{lee06,lee08}.
This needs to be tested in future studies.

\subsection{Spatial Distribution of Pixel Stellar Populations}

The spatial distributions of pixels in different color domains show how stellar populations vary according to their location.
In NGC 5194, there seem to be two important factors affecting the internal distribution of stellar populations.
The first is the effect of spiral arms. In Fig.~\ref{iring}, from the `before-arm' area to the `after-arm' area, it is clear that the fraction of the young population (P1) increases largely (by factor of $\sim1.9$ or 2.6) while the young and dusty population (P2) decreases (by factor of $\sim0.6$ or 0.8). These results are explained by the density wave model \citep{lin63,too77,ber89}, in which trailing spiral density waves proceed slower than rotating material, compressing gas and dust into new stars. The spatial distributions of pixels across the spiral arms of NGC 5194 are consistent with such a compressing process: dense dust (before-arm) $\rightarrow$ newly-formed stars (after-arm). The pixels highly-extinct by dust result in the plume features in the pCMD.

Another factor is the effect of the interaction with NGC 5195. While the pixels in the inner disk of NGC 5194 have relatively symmetric properties (that is, the trends along A1-A2-A3 are similar to those along A4-A5-A6), the pixels in the outer disk ($R>2$ arcmin = 4.8 kpc) show asymmetric properties in their population distribution. 
As shown in Fig.~\ref{oring}, the young population (P1) fraction in the NGC 5195 side (B4) is larger than that in the opposite side (B2) by factor of 2.3 and the trend along B3-B4 is clearly different from that along B1-B2.
This result indicates that the interaction with NGC 5195 significantly affects the stellar population in the area between NGC 5194 and NGC 5195, in the sense that the interaction enhances the star formation activity in the bridge area between NGC 5194 and NGC 5195.

\subsection{Optically Resolved AGN?}

\citet{lan07} reported hook features at the bright-end of the pCMDs in some early-type galaxies and explained them as the results of strong dust attenuation. However, the bright-end of the NGC 5194 pCMD has a reverse shape to those `red hooks' reported by \citet{lan07}; that is, brighter pixels have bluer colors in the bright-end part of the red pixel sequence.
The pixels forming the tight sequence at the bright-end of the pCMD are within $\sim100$ pc from the NGC 5194 center.

It is known that NGC 5194 hosts an AGN \citep{for85,ter01}, which is thought to significantly affect the photometric properties of the central area of NGC 5194.
Today, according to the AGN unified model \citep[e.g.][]{urr95}, AGNs with various properties are regarded as objects with the same structure but different viewing angles, which consist of a black hole, an accretion disk, an obscuring torus, jets, and so on.
Among those components of an AGN, the scale of an obscuring torus has been estimated with indirect methods (mainly from spectroscopic features) to be several parsecs \citep{pie92}, several tens of parsecs \citep{pie93}, or several hundreds of parsecs \citep{gra94,gra97}.

Although the estimates of the AGN torus size range somewhat widely, it is interesting that the size of the area occupied by the bright-end sequence pixels in the NGC 5194 pCMD is approximately consistent with the estimated size of the AGN torus. However, such agreement does not explain the origin of the tight sequence. If the light from those central pixels were stellar light, the obscuring torus would make those pixels redder, rather than bluer. Moreover, pixels closer to the NGC 5194 center tend to be bluer, showing a possibility that the bright-end sequence may be non-stellar light from the AGN itself.

If the bright-end sequence originates from the non-stellar light of AGN, the sequence features such as slope and color offset can be used as photometric parameters describing the properties of the AGN. Like many other pCMD parameters, this also needs to be tested using a large sample of AGNs, but this test requires very high-resolution images for sample AGNs. Since the approximate size of the bright-end pixel area is 100 pc in radius, spatial resolution better than a few tens of parsecs per pixel is necessary to estimate the sequence reasonably. If it is supposed that the acceptable resolution is 30 pc pixel$^{-1}$, the photometric pixel sequences of AGNs can be studied up to z $\sim0.015$ using the {\it HST} (0.1$''$ resolution) and up to z $\sim0.18$ using an adaptive optics system (0.01$''$ resolution).

\section{CONCLUSION}

We carried out a pixel analysis of the interacting face-on spiral galaxy NGC 5194, using {\it HST}/ACS images in the $BVI$ bands. Three major results were presented. First, we derived several quantities describing the pixel color-magnitude diagram (pCMD) features of NGC 5194, such as blue/red color cut, red pixel sequence parameters, blue pixel sequence parameters and blue-to-red pixel ratio. Those parameters reflect the internal properties of NGC 5194, such as age, metallicity, dust content and galaxy morphology.
In the luminosity-weighted mean stellar age estimation with solar metallicity and $\tau_V=1$ assumptions, the red sequence pixels are mostly older than 1 Gyr, while the blue sequence pixels are mostly younger than 1 Gyr.
The color variation in the red pixel sequence from $V=20$ mag arcsec$^{-2}$ to $V=17$ mag arcsec$^{-2}$ corresponds to a metallicity variation of $\Delta$[Fe/H] $\sim2$ or an optical depth variation of $\Delta\tau_V\sim4$ by dust, but the actual sequence is thought to originate from the combination of those two effects. At $V<20$ mag arcsec$^{-2}$, the color variation in the blue pixel sequence corresponds to age variation from 5 Myr to 300 Myr in solar metallicity and $\tau_V=1$ assumptions.
The spatial resolution dependence of the pCMD features was tested for the future comparisons between galaxies with different image resolutions, and we found that the pixel resolution needs to be better than 100 pc pixel$^{-1}$ for the investigation of pCMD features.

Second, we found that the spatial distributions of pixel stellar populations are significantly affected by the spiral arm patterns and the tidal interaction with NGC 5195. The pixel population distributions across the spiral arms agree with a compressing process by spiral density waves: dense dust (before-arm) $\rightarrow$ newly-formed stars (after-arm).
The interaction between NGC 5194 and NGC 5195 also enhances the star formation at the tidal bridge area connecting the two galaxies.

Finally, we found that the pixels corresponding to the central area (around the AGN) of NGC 5194 show a tight sequence in the bright-end of the pCMD, of which spatial extent is $R\sim100$ parsecs. The origin and mechanism of the tight bright-end pixel sequence are not sufficiently explained in this paper, but the sequence seems to be worth a comparison between various AGNs using high-resolution imaging data in the future, because it may be a photometric indicator of AGN properties.

\acknowledgments

This paper presents the first results of the pilot studies for an observational research project, Pixel Analysis of Nearby Cluster Galaxies (PANCluG).
All authors are the members of Dedicated Researchers for Extragalactic AstronoMy (DREAM) in Korea Astronomy and Space Science Institute (KASI). H.S.P.~was supported by Mid-career Researcher Program through NRF grant funded by the MEST (No.2010-0013875).
We appreciate the anonymous referee for the useful comments.

\begin{deluxetable}{crrrrrrrrr}
\tablenum{1} \tablecolumns{10} \tablecaption{ Gaussian Fit Parameters of the Color Distributions as a Function of Pixel Surface Brightness\label{gauss}} \tablewidth{0pt}
\tablehead{$V$ range  & \multicolumn{3}{c}{Gaussian 1} & \multicolumn{3}{c}{Gaussian 2}  & \multicolumn{3}{c}{Gaussian 3}\\
(mag arcsec$^{-2}$) & $A_1$ & $x_1$ & $\sigma_1$ & $A_2$ & $x_2$ & $\sigma_2$ & $A_3$ & $x_3$ & $\sigma_3$}
\startdata
\multicolumn{10}{c}{(For $B-V$)} \\
17.5 -- 18.0 & 0.664 & 0.749 & 0.050 & --- & --- & --- & --- & --- & --- \\
18.0 -- 18.5 & 0.503 & 0.489 & 0.183 & 0.843 & 0.627 & 0.0601 & 0.844 & 0.763 & 0.033 \\
18.5 -- 19.0 & 0.375 & 0.275 & 0.245 & 1.313 & 0.585 & 0.133 & --- & --- & --- \\
19.0 -- 19.5 & 0.870 & 0.226 & 0.227 & 1.942 & 0.555 & 0.136 & --- & --- & --- \\
19.5 -- 20.0 & 4.615 & 0.345 & 0.203 & 1.865 & 0.611 & 0.066 & --- & --- & --- \\
20.0 -- 20.5 & 26.665 & 0.426 & 0.155 & 10.555 & 0.638 & 0.047 & --- & --- & --- \\
20.5 -- 21.0 & 96.411 & 0.482 & 0.143 & --- & --- & --- & --- & --- & --- \\
&&&&&&\\
18.5 -- 19.5 & 1.255 & 0.247 & 0.237 & 3.199 & 0.568 & 0.135 & --- & --- & --- \\
\hline
\\
\multicolumn{10}{c}{(For $V-I$)} \\
17.5 -- 18.0 & 0.463 & 0.803 & 0.080 & --- & --- & --- & --- & --- & --- \\
18.0 -- 18.5 & 1.056 & 0.730 & 0.136 & 0.326 & 0.887 & 0.048 & --- & --- & --- \\
18.5 -- 19.0 & 0.202 & 0.238 & 0.327 & 1.314 & 0.748 & 0.155 & --- & --- & --- \\
19.0 -- 19.5 & 0.527 & 0.183 & 0.243 & 2.024 & 0.724 & 0.165 & --- & --- & --- \\
19.5 -- 20.0 & 3.448 & 0.481 & 0.259 & 1.422 & 0.813 & 0.126 & --- & --- & --- \\
20.0 -- 20.5 & 23.941 & 0.591 & 0.197 & --- & --- & --- & --- & --- & --- \\
20.5 -- 21.0 & 78.954 & 0.601 & 0.175 & --- & --- & --- & --- & --- & --- \\
&&&&&&\\
18.5 -- 19.5 & 0.723 & 0.212 & 0.279 & 3.310 & 0.736 & 0.160 & --- & --- & --- \\
\enddata
\tablecomments{ The fitting function formula is as follows:\\
Fit($x$) $= A_1 \cdot \exp(-(x-x_1)^2/\,2\,{\sigma_1}^2) + A_2 \cdot \exp(-(x-x_2)^2/\,2\,{\sigma_2}^2) + A_3 \cdot \exp(-(x-x_3)^2/\,2\,{\sigma_3}^2)$, \\
where $x$ is the color index ($B-V$ or $V-I$).
}
\end{deluxetable}

\begin{deluxetable}{ccccc}
\tablenum{2} \tablecolumns{8} \tablecaption{ Several Parameters of the NGC 5194 pCMD ($4\times4$ Binning) \label{pcmdinfo}} \tablewidth{0pt}
\tablehead{ $V$ range & Total pixels & Blue pixels & Red pixels & Blue/red ratio$^{(1)}$}
\startdata
& & $B-V$ (cut $=0.299$) & & \\
& Median (SIQR) & Median (SIQR) & Median (SIQR) & \\
\hline
16.0 -- 16.5 & 0.534 (0.299) & $-0.032$ (0.092)~~ & 0.624 (0.158) & $0.473\pm0.113$ \\
16.5 -- 17.0 & 0.698 (0.081) & $-0.022$ (0.122)~~ & 0.715 (0.030) & $0.303\pm0.055$ \\
17.0 -- 17.5 & 0.747 (0.191) & 0.001 (0.124) & 0.762 (0.028) & $0.291\pm0.028$ \\
17.5 -- 18.0 & 0.718 (0.115) & 0.064 (0.139) & 0.735 (0.054) & $0.191\pm0.011$ \\
18.0 -- 18.5 & 0.593 (0.126) & 0.121 (0.131) & 0.617 (0.099) & $0.143\pm0.005$ \\
18.5 -- 19.0 & 0.515 (0.141) & 0.115 (0.123) & 0.564 (0.100) & $0.248\pm0.005$ \\
19.0 -- 19.5 & 0.452 (0.163) & 0.115 (0.109) & 0.532 (0.100) & $0.392\pm0.006$ \\
19.5 -- 20.0 & 0.370 (0.156) & 0.162 (0.091) & 0.490 (0.108) & $0.561\pm0.005$ \\
20.0 -- 20.5 & 0.440 (0.120) & 0.201 (0.069) & 0.487 (0.099) & $0.255\pm0.001$ \\
20.5 -- 21.0 & 0.477 (0.098) & 0.235 (0.048) & 0.501 (0.082) & $0.149\pm0.001$ \\
21.0 -- 21.5 & 0.451 (0.073) & 0.250 (0.038) & 0.464 (0.065) & $0.106\pm0.000$ \\
21.5 -- 22.0 & 0.461 (0.069) & 0.252 (0.039) & 0.470 (0.063) & $0.081\pm0.000$ \\
22.0 -- 22.5 & 0.489 (0.069) & 0.261 (0.031) & 0.494 (0.067) & $0.038\pm0.000$ \\
22.5 -- 23.0 & 0.521 (0.076) & 0.264 (0.027) & 0.525 (0.073) & $0.027\pm0.000$ \\
23.0 -- 23.5 & 0.283 (0.070) & 0.222 (0.053) & 0.362 (0.040) & $1.330\pm0.044$ \\
\hline
&&&&\\
& & $V-I$ (cut $=0.416$) & & \\
& Median (SIQR) & Median (SIQR) & Median (SIQR) &\\
\hline
16.0 -- 16.5 & 0.638 (0.344) & $-0.057$ (0.228)~~ & 0.725 (0.039) & $0.833\pm0.191$ \\
16.5 -- 17.0 & 0.767 (0.229) & $-0.057$ (0.195)~~ & 0.785 (0.044) & $0.352\pm0.062$ \\
17.0 -- 17.5 & 0.793 (0.289) & 0.008 (0.216) & 0.830 (0.050) & $0.395\pm0.035$ \\
17.5 -- 18.0 & 0.774 (0.104) & 0.047 (0.195) & 0.794 (0.059) & $0.206\pm0.011$ \\
18.0 -- 18.5 & 0.721 (0.119) & 0.103 (0.184) & 0.746 (0.096) & $0.138\pm0.004$ \\
18.5 -- 19.0 & 0.688 (0.152) & 0.134 (0.175) & 0.737 (0.108) & $0.227\pm0.005$ \\
19.0 -- 19.5 & 0.644 (0.188) & 0.156 (0.144) & 0.719 (0.112) & $0.337\pm0.005$ \\
19.5 -- 20.0 & 0.536 (0.194) & 0.257 (0.112) & 0.673 (0.134) & $0.513\pm0.005$ \\
20.0 -- 20.5 & 0.584 (0.134) & 0.314 (0.076) & 0.638 (0.108) & $0.245\pm0.001$ \\
20.5 -- 21.0 & 0.599 (0.118) & 0.346 (0.051) & 0.631 (0.102) & $0.168\pm0.001$ \\
21.0 -- 21.5 & 0.533 (0.095) & 0.360 (0.040) & 0.567 (0.080) & $0.235\pm0.001$ \\
21.5 -- 22.0 & 0.506 (0.085) & 0.367 (0.034) & 0.542 (0.071) & $0.284\pm0.001$ \\
22.0 -- 22.5 & 0.480 (0.071) & 0.371 (0.031) & 0.515 (0.059) & $0.346\pm0.001$ \\
22.5 -- 23.0 & 0.470 (0.064) & 0.371 (0.030) & 0.504 (0.051) & $0.388\pm0.001$ \\
23.0 -- 23.5 & 0.641 (0.055) & 0.373 (0.023) & 0.642 (0.055) & $0.003\pm0.001$ \\
\enddata
\tablecomments{(1) Pixel number ratio $\pm$ Poisson error.}
\end{deluxetable}

\begin{deluxetable}{cccccc}
\tablenum{3} \tablecolumns{6} \tablecaption{ $(B-V)$ pCMD Variation along Spatial Resolution \label{pcmdvar1}} \tablewidth{0pt}
\tablehead{ (1) & ~~~~$4\times4$ $^{(3)}$ & $20\times20$ & $50\times50$ & $100\times100$ & $200\times200$ \\
(2) & 8 & 40 & 100 & 200 & 400 }
\startdata
$V$ range & \multicolumn{5}{c}{Blue pixel sequence} \\
\hline
16.0 -- 16.5 & $-0.331$ (0.092) & ~~~---$^{(4)}$ & --- & --- & --- \\
16.5 -- 17.0 & $-0.321$ (0.122) & --- & --- & --- & --- \\
17.0 -- 17.5 & $-0.298$ (0.123) & --- & --- & --- & --- \\
17.5 -- 18.0 & $-0.235$ (0.139) & $-0.122$ (0.048) & --- & --- & --- \\
18.0 -- 18.5 & $-0.178$ (0.131) & $-0.234$ (0.142) & --- & --- & --- \\
18.5 -- 19.0 & $-0.184$ (0.123) & $-0.189$ (0.102) & $-0.184$ (0.082) & --- & --- \\
19.0 -- 19.5 & $-0.184$ (0.109) & $-0.160$ (0.088) & $-0.132$ (0.077) & $-0.137$ (0.025) & --- \\
19.5 -- 20.0 & $-0.137$ (0.091) & $-0.128$ (0.077) & $-0.139$ (0.072) & $-0.096$ (0.060) & $-0.007$ (0.006) \\
20.0 -- 20.5 & $-0.098$ (0.069) & $-0.084$ (0.051) & $-0.089$ (0.047) & $-0.080$ (0.050) & $-0.109$ (0.038) \\
20.5 -- 21.0 & $-0.064$ (0.048) & $-0.047$ (0.034) & $-0.038$ (0.030) & $-0.043$ (0.029) & $-0.032$ (0.022) \\
21.0 -- 21.5 & $-0.049$ (0.038) & $-0.035$ (0.026) & $-0.032$ (0.024) & $-0.031$ (0.027) & $-0.035$ (0.018) \\
21.5 -- 22.0 & $-0.047$ (0.039) & $-0.029$ (0.023) & $-0.023$ (0.017) & $-0.013$ (0.009) & $-0.015$ (0.014) \\
22.0 -- 22.5 & $-0.038$ (0.031) & $-0.024$ (0.020) & $-0.028$ (0.010) & $-0.006$ (0.003) & --- \\
22.5 -- 23.0 & $-0.035$ (0.027) & $-0.024$ (0.009) & --- & --- & --- \\
23.0 -- 23.5 & $-0.077$ (0.053) & --- & --- & --- & --- \\
\hline\hline
$V$ range & \multicolumn{5}{c}{Red pixel sequence} \\
\hline
16.0 -- 16.5 & 0.325 (0.157) & --- & --- & --- & --- \\ 
16.5 -- 17.0 & 0.415 (0.030) & 0.432 (0.018) & --- & --- & --- \\ 
17.0 -- 17.5 & 0.463 (0.028) & 0.459 (0.020) & 0.451 (0.011) & --- & --- \\ 
17.5 -- 18.0 & 0.436 (0.054) & 0.451 (0.028) & 0.453 (0.013) & 0.450 (0.028) & --- \\ 
18.0 -- 18.5 & 0.318 (0.099) & 0.310 (0.083) & 0.329 (0.068) & 0.339 (0.052) & 0.355 (0.035) \\ 
18.5 -- 19.0 & 0.264 (0.099) & 0.232 (0.088) & 0.211 (0.064) & 0.243 (0.031) & 0.235 (0.026) \\ 
19.0 -- 19.5 & 0.233 (0.100) & 0.206 (0.083) & 0.188 (0.077) & 0.178 (0.042) & 0.204 (0.050) \\ 
19.5 -- 20.0 & 0.191 (0.108) & 0.170 (0.104) & 0.173 (0.093) & 0.162 (0.079) & 0.153 (0.065) \\ 
20.0 -- 20.5 & 0.188 (0.099) & 0.168 (0.099) & 0.162 (0.099) & 0.168 (0.102) & 0.145 (0.098) \\ 
20.5 -- 21.0 & 0.202 (0.082) & 0.185 (0.079) & 0.181 (0.075) & 0.173 (0.072) & 0.168 (0.071) \\ 
21.0 -- 21.5 & 0.164 (0.065) & 0.145 (0.056) & 0.137 (0.053) & 0.135 (0.048) & 0.130 (0.047) \\ 
21.5 -- 22.0 & 0.171 (0.062) & 0.154 (0.052) & 0.148 (0.050) & 0.147 (0.050) & 0.145 (0.046) \\ 
22.0 -- 22.5 & 0.195 (0.067) & 0.184 (0.039) & 0.182 (0.035) & 0.179 (0.036) & 0.179 (0.039) \\ 
22.5 -- 23.0 & 0.226 (0.073) & 0.240 (0.046) & 0.250 (0.047) & 0.255 (0.047) & 0.260 (0.048) \\ 
23.0 -- 23.5 & 0.063 (0.040) & --- & --- & --- & --- \\
\enddata
\tablecomments{ (1) Binning factor. (2) Pixel scale in unit of pc pixel$^{-1}$. (3) Median color $-$ 0.299 (SIQR; sample inter-quartile range). (4) No or single pixel. }
\end{deluxetable}

\begin{deluxetable}{cccccc}
\tablenum{4} \tablecolumns{6} \tablecaption{ $(V-I)$ pCMD Variation along Spatial Resolution \label{pcmdvar2}} \tablewidth{0pt}
\tablehead{ (1) & ~~~~$4\times4$ $^{(3)}$ & $20\times20$ & $50\times50$ & $100\times100$ & $200\times200$ \\
(2) & 8 & 40 & 100 & 200 & 400 }
\startdata
$V$ range & \multicolumn{5}{c}{Blue pixel sequence} \\
\hline
16.0 -- 16.5 & $-0.473$ (0.228) & ~~~---$^{(4)}$ & --- & --- & --- \\
16.5 -- 17.0 & $-0.473$ (0.194) & --- & --- & --- & --- \\
17.0 -- 17.5 & $-0.408$ (0.216) & $-0.581$ (0.010) & --- & --- & --- \\
17.5 -- 18.0 & $-0.369$ (0.195) & $-0.279$ (0.236) & --- & --- & --- \\
18.0 -- 18.5 & $-0.313$ (0.184) & $-0.389$ (0.155) & --- & --- & --- \\
18.5 -- 19.0 & $-0.282$ (0.175) & $-0.267$ (0.116) & $-0.278$ (0.153) & --- & --- \\
19.0 -- 19.5 & $-0.260$ (0.144) & $-0.175$ (0.107) & $-0.176$ (0.050) & $-0.079$ (0.106) & --- \\
19.5 -- 20.0 & $-0.159$ (0.112) & $-0.126$ (0.079) & $-0.124$ (0.066) & $-0.056$ (0.066) & --- \\
20.0 -- 20.5 & $-0.102$ (0.076) & $-0.078$ (0.052) & $-0.082$ (0.044) & $-0.073$ (0.040) & $-0.081$ (0.034) \\
20.5 -- 21.0 & $-0.070$ (0.051) & $-0.048$ (0.036) & $-0.040$ (0.029) & $-0.038$ (0.030) & $-0.034$ (0.019) \\
21.0 -- 21.5 & $-0.056$ (0.040) & $-0.039$ (0.028) & $-0.034$ (0.023) & $-0.034$ (0.022) & $-0.023$ (0.016) \\
21.5 -- 22.0 & $-0.049$ (0.034) & $-0.030$ (0.021) & $-0.026$ (0.018) & $-0.026$ (0.015) & $-0.022$ (0.011) \\
22.0 -- 22.5 & $-0.045$ (0.031) & $-0.019$ (0.015) & $-0.019$ (0.011) & $-0.018$ (0.011) & $-0.009$ (0.008) \\
22.5 -- 23.0 & $-0.045$ (0.030) & $-0.014$ (0.011) & $-0.014$ (0.008) & $-0.013$ (0.008) & $-0.015$ (0.003) \\
23.0 -- 23.5 & $-0.043$ (0.023) & --- & --- & --- & --- \\
\hline\hline
$V$ range & \multicolumn{5}{c}{Red pixel sequence} \\
\hline
16.0 -- 16.5 & 0.309 (0.039) & --- & --- & --- & --- \\ 
16.5 -- 17.0 & 0.369 (0.044) & 0.431 (0.027) & --- & --- & --- \\ 
17.0 -- 17.5 & 0.414 (0.050) & 0.425 (0.029) & 0.385 (0.049) & --- & --- \\ 
17.5 -- 18.0 & 0.378 (0.059) & 0.401 (0.040) & 0.420 (0.025) & 0.431 (0.029) & --- \\ 
18.0 -- 18.5 & 0.330 (0.096) & 0.342 (0.069) & 0.361 (0.047) & 0.359 (0.035) & 0.381 (0.027) \\ 
18.5 -- 19.0 & 0.321 (0.108) & 0.302 (0.083) & 0.301 (0.058) & 0.342 (0.038) & 0.327 (0.024) \\ 
19.0 -- 19.5 & 0.302 (0.112) & 0.283 (0.092) & 0.277 (0.075) & 0.291 (0.035) & 0.314 (0.030) \\ 
19.5 -- 20.0 & 0.257 (0.134) & 0.241 (0.132) & 0.253 (0.118) & 0.256 (0.090) & 0.278 (0.093) \\ 
20.0 -- 20.5 & 0.222 (0.108) & 0.207 (0.108) & 0.207 (0.107) & 0.219 (0.111) & 0.221 (0.104) \\ 
20.5 -- 21.0 & 0.215 (0.102) & 0.198 (0.094) & 0.194 (0.089) & 0.190 (0.084) & 0.177 (0.084) \\ 
21.0 -- 21.5 & 0.151 (0.080) & 0.129 (0.067) & 0.119 (0.061) & 0.115 (0.056) & 0.110 (0.049) \\ 
21.5 -- 22.0 & 0.126 (0.071) & 0.101 (0.058) & 0.090 (0.056) & 0.086 (0.054) & 0.081 (0.052) \\ 
22.0 -- 22.5 & 0.099 (0.059) & 0.078 (0.038) & 0.073 (0.032) & 0.070 (0.029) & 0.068 (0.031) \\ 
22.5 -- 23.0 & 0.088 (0.051) & 0.057 (0.024) & 0.056 (0.018) & 0.056 (0.016) & 0.058 (0.015) \\ 
23.0 -- 23.5 & 0.226 (0.055) & --- & --- & --- & --- \\
\enddata
\tablecomments{ (1) Binning factor. (2) Pixel scale in unit of pc pixel$^{-1}$. (3) Median color $-$ 0.416 (SIQR). (4) No or single pixel. }
\end{deluxetable}

\begin{deluxetable}{cccccc}
\tablenum{5} \tablecolumns{6} \tablecaption{ Blue/Red Pixel Number Ratio $\pm$ Poisson Error along Spatial Resolution \label{pcmdvar3}} \tablewidth{0pt}
\tablehead{ (1) & $4\times4$ & $20\times20$ & $50\times50$ & $100\times100$ & $200\times200$ \\
(2) & 8 & 40 & 100 & 200 & 400 }
\startdata
$V$ range & \multicolumn{5}{c}{$(B-V)$ pCMD} \\
\hline
16.0 -- 16.5 & $0.473\pm0.113$ & ~~~---$^{(3)}$ & --- & --- & --- \\
16.5 -- 17.0 & $0.303\pm0.055$ & --- & --- & --- & --- \\
17.0 -- 17.5 & $0.291\pm0.028$ & --- & --- & --- & --- \\
17.5 -- 18.0 & $0.191\pm0.011$ & $0.081\pm0.034$ & --- & --- & --- \\
18.0 -- 18.5 & $0.143\pm0.005$ & $0.089\pm0.018$ & --- & --- & --- \\
18.5 -- 19.0 & $0.248\pm0.005$ & $0.174\pm0.022$ & $0.118\pm0.042$ & --- & --- \\
19.0 -- 19.5 & $0.392\pm0.006$ & $0.352\pm0.027$ & $0.278\pm0.056$ & $0.161\pm0.078$ & --- \\
19.5 -- 20.0 & $0.561\pm0.005$ & $0.593\pm0.027$ & $0.532\pm0.061$ & $0.368\pm0.094$ & $0.133\pm0.100$ \\
20.0 -- 20.5 & $0.255\pm0.001$ & $0.253\pm0.006$ & $0.232\pm0.015$ & $0.257\pm0.033$ & $0.202\pm0.054$ \\
20.5 -- 21.0 & $0.149\pm0.001$ & $0.143\pm0.003$ & $0.137\pm0.006$ & $0.130\pm0.012$ & $0.152\pm0.026$ \\
21.0 -- 21.5 & $0.106\pm0.000$ & $0.082\pm0.002$ & $0.079\pm0.004$ & $0.078\pm0.008$ & $0.065\pm0.014$ \\
21.5 -- 22.0 & $0.081\pm0.000$ & $0.036\pm0.001$ & $0.031\pm0.003$ & $0.029\pm0.005$ & $0.021\pm0.009$ \\
22.0 -- 22.5 & $0.038\pm0.000$ & $0.005\pm0.000$ & $0.003\pm0.001$ & $0.003\pm0.002$ & --- \\
22.5 -- 23.0 & $0.027\pm0.000$ & $0.000\pm0.000$ & --- & --- & --- \\
23.0 -- 23.5 & $1.330\pm0.044$ & --- & --- & --- & --- \\
\hline\hline
$V$ range & \multicolumn{5}{c}{$(V-I)$ pCMD} \\
\hline
16.0 -- 16.5 & $0.833\pm0.191$ & --- & --- & --- & --- \\
16.5 -- 17.0 & $0.352\pm0.062$ & --- & --- & --- & --- \\
17.0 -- 17.5 & $0.395\pm0.035$ & $0.111\pm0.083$ & --- & --- & --- \\
17.5 -- 18.0 & $0.206\pm0.011$ & $0.099\pm0.039$ & --- & --- & --- \\
18.0 -- 18.5 & $0.138\pm0.004$ & $0.068\pm0.015$ & --- & --- & --- \\
18.5 -- 19.0 & $0.227\pm0.005$ & $0.129\pm0.018$ & $0.104\pm0.039$ & --- & --- \\
19.0 -- 19.5 & $0.337\pm0.005$ & $0.241\pm0.021$ & $0.167\pm0.039$ & $0.091\pm0.055$ & --- \\
19.5 -- 20.0 & $0.513\pm0.005$ & $0.447\pm0.021$ & $0.380\pm0.046$ & $0.300\pm0.081$ & --- \\
20.0 -- 20.5 & $0.245\pm0.001$ & $0.198\pm0.005$ & $0.182\pm0.013$ & $0.163\pm0.024$ & $0.161\pm0.046$ \\
20.5 -- 21.0 & $0.168\pm0.001$ & $0.135\pm0.002$ & $0.117\pm0.006$ & $0.107\pm0.010$ & $0.082\pm0.018$ \\
21.0 -- 21.5 & $0.235\pm0.001$ & $0.165\pm0.002$ & $0.151\pm0.006$ & $0.140\pm0.011$ & $0.134\pm0.022$ \\
21.5 -- 22.0 & $0.284\pm0.001$ & $0.159\pm0.003$ & $0.135\pm0.006$ & $0.116\pm0.011$ & $0.120\pm0.023$ \\
22.0 -- 22.5 & $0.346\pm0.001$ & $0.099\pm0.002$ & $0.063\pm0.004$ & $0.061\pm0.008$ & $0.055\pm0.015$ \\
22.5 -- 23.0 & $0.388\pm0.001$ & $0.083\pm0.002$ & $0.045\pm0.003$ & $0.037\pm0.006$ & $0.030\pm0.012$ \\
23.0 -- 23.5 & $0.003\pm0.001$ & --- & --- & --- & --- \\
\enddata
\tablecomments{ (1) Binning factor. (2) Pixel scale in unit of pc pixel$^{-1}$. (3) No or single pixel. }
\end{deluxetable}

\end{document}